\documentclass[10pt]{article}
\usepackage[utf8]{inputenc}
\usepackage{amsmath}
\usepackage{authblk}
\usepackage{hyperref}
\usepackage{soul}
\usepackage{color, xcolor}
\usepackage{graphicx}
\graphicspath{{}}
\usepackage{geometry}
\usepackage{booktabs}
\usepackage{array}
\usepackage{float}
\newcommand{\myref}[1]{equation(s) \ref{#1}}
\newcommand{\fref}[1]{figure \ref{#1}}
\newcommand{\Fref}[1]{Figure \ref{#1}}
\newcommand{\pref}[1]{proposition \ref{#1}}

\usepackage{caption}
\usepackage[title]{appendix}
\usepackage{ntheorem}
\theoremseparator{.}
\newtheorem{proposition}{Proposition.}
\usepackage{indentfirst}

\title{Filtered Partial Differential Equations: a robust surrogate constraint in physics-informed deep learning framework}

\author[a]{Dashan Zhang}
\author[b]{Yuntian Chen \thanks{Corresponding author: ychen@eitech.edu.cn}}
\author[b, a]{Shiyi Chen \thanks{Corresponding author: chensy@sustech.edu.cn}}
\affil[a]{Department of Mechanics and Aerospace Engineering, Southern University of Science and Technology, Shenzhen 518055, China}
\affil[b]{Ningbo Institute of Digital Twin, Eastern Institute of Technology, Ningbo 315200, China}

\date{}

\begin{document}

% \linenumbers
% \pagewiselinenumbers

\newgeometry{left=2.8cm,right=2.8cm,top=2.5cm,bottom=2.5cm}

\restoregeometry{}

\maketitle

\begin{abstract}
Embedding physical knowledge into neural network (NN) training has been a hot topic. However, when facing the complex real-world, most of the existing methods still strongly rely on the quantity and quality of observation data. Furthermore, the neural networks often struggle to converge when the solution to the real equation is very complex. Inspired by large eddy simulation (LES) in computational fluid dynamics, we propose an improved method based on filtering. We analyzed the causes of the difficulties in physics informed machine learning, and proposed a surrogate constraint (filtered PDE, FPDE in short) of the original physical equations to reduce the influence of noisy and sparse observation data. In the noise and sparsity experiment, the proposed FPDE models (which are optimized by FPDE constraints) have better robustness than the conventional PDE models. Experiments demonstrate that the FPDE model can obtain the same quality solution with 100\% higher noise and 12\% quantity of observation data of the baseline. Besides, two groups of real measurement data are used to show the FPDE improvements in real cases. The final results show that FPDE still gives more physically reasonable solutions when facing the incomplete equation problem and the extremely sparse and high-noise conditions. The proposed FPDE constraint is helpful for merging real-world experiment data into physics-informed training, and it works effectively in two real-world experiments: simulating cell movement in scratches and blood velocity in vessels.

%Inspired by the numerical method in Computational Fluid Dynamics (CFD), we raised Filtered Partial Differential Equations (FPDE) constraint, a surrogate constraint of the residual form Partial Differential Equations (PDE) constraint. The proposed FPDE optimized physics-informed Neural network (PINN) model have better robustness than the PDE constrained model when facing the inaccurate data. The comparison shows the performance of FPDE model less dependent on data quality than the original model, which has reduced L2 error by more than 80\% under sparse data and 70\% under noisy data. Analyzed the reason for the difficulty in solving complex PDE problems by NN, we used the concept of ‘Conflict’ to explain the existence of local optimum in the co-optimization process. Firstly, we use the simulation data with artificial noise and sparse sampling to show the FPDE improvements in different quality and quantity of data. Then, the real measurement data of two experiments (cell migration and arterial flow) is used to verify that the proposed FPDE performs well in real-world problems. Under the data with sparse and high noise, mismatch with theoretical equation and even unknown coefficient, FPDE always can make improvements and give reasonable solutions.  In the last part, the existence of ‘Conflict’ is found in the analysis of training process, which further validated our improvements.
\end{abstract}


\section{Introduction}
\label{Introduction}

% Deep learning is one of the most powerful methods for fitting an unknown data distribution. In this section, a brief introduction of the physics-informed framework is made as the background knowledge. 
One of the most effective techniques for fitting an unknown data distribution is deep learning. As background information, a brief introduction to the physics-informed framework is provided in this section. This framework utilizes the governing equation to better solve problems and increase the model's performance. Meanwhile, we demonstrate the challenge inherent in this framework with some examples. Due to the complexity of the original equations, the optimization of the neural network becomes difficult. Therefore, the idea of simplifying the issue via surrogate equations is brought forward. Correspondingly, this improvement can lessen the reliance on the quality and quantity of data in physics-informed training. 

\subsection {Background of solving PDE via NN}
 Since  the advent of scientific inquiry, scholars have endeavored to formulate and resolve equations to elucidate natural phenomena. Differential equations, which contain derivatives of unknown variables, represent a cornerstone of both physical and mathematical discourse. For a large number of partial differential equation (PDE) systems, it is challenging to directly obtain their analytical solutions. Therefore, numerous numerical methods have been developed and employed to approximate the solutions of PDEs through simulation.
 % The ordinary differential equation (ODE) has a unique solution when the Lipschitz continuity condition is satisfied \cite{kirszbraun1934zusammenziehende}, but the partial differential equation (PDE) doesn't have such good properties. Therefore, many numerical methods have been designed and used to find the approximate solutions of PDEs by simulation.
 
With the development of artificial intelligence, data-driven models are widely used in many disciplines  \cite{littmann2020validity, debroy2021metallurgy, goodell2021artificial, wang2023ai}. The Neural network (NN) models show their strong fitting ability in Computer vision (CV, \cite{vernon1991machine}) and Natural language processing (NLP, \cite{sag2002multiword}) field. In the field of engineering computation, the Fourier neural operator (FNO) is proposed to learn the features in Spectral space \cite{li2020fourier}; deconvolutional artificial neural network (DANN) is developed for subgrid-scale (SGS) stress in large eddy simulation (LES) of turbulence \cite{yuan2020deconvolutional}; generative adversarial network (GAN) is also used to generate complex turbulence under the condition of missing data due to its good fidelity \cite{li2022data}. The inherent abstract reasoning process of neural networks (NNs) empowers them to adeptly learn embedding mappings across a wide array of training datasets. With the gradient descent \cite{ruder2016overview} method, the NN can learn the pattern, which is the relationship of different features from the data in the optimization process. Many kinds of NN architecture are raised, such as the convolutional block \cite{fukushima1982neocognitron} and the self-attention block \cite{zhang2019self}, to treat different features. But when the data is insufficient to cover the features in embedding space, the question arises: can domain knowledge be incorporated to enhance the optimization process?

The paradigms that add equation constraints into the optimization process of NN, such as the physics-informed neural network (PINN, \cite{raissi2019physics}), provide a beautiful vision of solving PDE with domain knowledge automatically. Training models with explicit physical constraints, such as the incorporation of governing equations, has shown promise in yielding improved results. Over the past four years, there has been a notable surge in the utilization of physics-informed methods across diverse domains. In the physics-informed based problem, some scientific machine learning frameworks are proposed (e.g., the DeepXDE, \cite{lu2021deepxde}; AutoKE, \cite{du2022autoke}; the NeuroDiffEq, \cite{chen2020neurodiffeq} framework) to solve the differential equations. In the inverse problem of physics-informed learning, the model is built for the equation discovering. Scholars have proposed many methods that can work in the knowledge discovery field \cite{chen2022integration}, the sparse regression method is capable of discovering the PDEs in given system \cite{rudy2017data}; the deep learning has also been proved to be effective on the physics-informed inverse problem (e.g., the DL-PDE, \cite{xu2019dl}); the symbolic genetic algorithm (SGA-PDE) can be used to discover the open-form PDEs \cite{chen2021any}; based on the Reynolds-averaged Navier–Stokes (RANS) equations, the physics-informed model is used to improve turbulence models \cite{duraisamy2019turbulence}. In the evaluation of PINN, PINN has been shown to be robust against the influence of sparsity and noise levels in training data \cite{clark2023reconstructing}, but its accuracy diminishes beyond the training time horizon \cite{du2023state}. In short, physics-informed model is generally considered as a modeling tool, particularly in contexts such as turbulence modeling, which have been demonstrated in isolated scenarios \cite{duraisamy2021perspectives}.

In the view of application, physics-informed framework has improved the model performance in many scenarios: using the physical laws in power systems, NN can model the power system behavior both in steady-state and in dynamics \cite{misyris2020physics}; the theory-guided deep-learning load forecasting (TgDLF) model the future load through load ratio decomposition with the considered historical load, weather forecast and calendar effect \cite{chen2021theory}; digital twins have been widely mentioned as an important preface application, highlighting the significance of modeling with physical constraints for their implementation \cite{rasheed2020digital}; in the subgrid modelling of Kraichnan turbulence, the data-driven method can predict the turbulence source term through localized grid-resolved information \cite{maulik2019subgrid}; with acoustic wave equation, NN can identify and characterize a surface breaking crack in a metal plate \cite{shukla2020physics}; with the advection-diffusion equations, NN can obtain better super-resolution outputs in the images of atmospheric pollution \cite{wang2020physics}; through the PINN-based method, three-dimensional Tomographic Background Oriented Schlieren imaging fields, such as the temperature field of an espresso cup, can be rapidly modeled \cite{cai2021flow}; using two-dimensional three-component stereo particle-image velocimetry (PIV) datasets, PINN demonstrates the capability to reconstruct the mean velocity field and correct measurement errors (\cite{wang2022dense}, \cite{hasanuzzaman2023enhancement}); based on stochastic particle advection velocimetry (SPAV) data, the PINN approach significantly improves the accuracy of particle tracking velocimetry reconstructions \cite{zhou2023stochastic}; in the field of geophysics, NN have shown enhanced ability to model the subsurface flow with the guiding of theory (e.g., governing equations, engineering controls and expert knowledge, \cite{wang2020deep}).

Unlike the numerical simulation method, the universal approximation theorem shows that a NN is able to approximate any continuous function on a compact subset of $ \mathcal{R}_n $ with sufficient precision when it has enough parameters \cite{hartman1990layered}. Considering that the optimization process of NN can be summarized as ‘finding the parameters to minimize the given loss function’, the equations can be added into the loss function to make NN solve PDE automatically. When facing a complex problem, adding some simulation or measurement data points to help NN determine the large-scale distribution of solutions is a common method. Subsequently, we refer to these data points as ‘observation points’.
%Therefore, it is theoretically possible to find the solution of any equation only with PDE constraints as long as the equations are closed.

Regarding solving equations, the Navier-Stokes \cite{temam1995navier} equation (\myref{NS classic}) in fluid mechanics is one of the most challenging problems. Both exact solutions \cite{wang1991} and computer simulations \cite{glowinski1992finite} play important roles in theory and engineering. When it comes to physics-informed training in fluid fields, modeling with the constraint of N-S equations is one of the most common methods. 
\begin{equation} \label{NS classic}
    \frac{\partial u}{\partial t}+u \cdot \nabla u = -\frac{1}{\rho}\nabla p + \nu \nabla^2u
\end{equation}
\begin{equation} \label{Continuity}
    \nabla \cdot u = 0
\end{equation}
where $u, p, \rho , \nu$ represent the velocity vector, pressure, density and dynamic viscosity, respectively.

\begin{figure}
    \centering
    \includegraphics[width=0.5\textwidth]{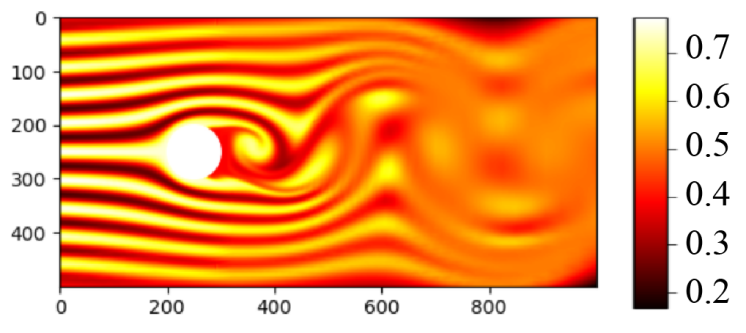}
    \caption{The fluid flow through a cylinder wall (streak-lines). To keep the consistency of data, the reference simulation solutions are from the open-resource data of Hidden fluid mechanics, HFM \cite{raissi2020hidden}}
    \label{fig:c_cylinder}
\end{figure}

In this paper, we utilize cylinder flow as an illustrative example to elucidate the challenges encountered in physics-informed training. We introduce our enhanced methodology and validate its effectiveness across diverse scenarios. The cylinder flow, governed by the N-S equation, is a classic example which can reflect the properties of the fluid \cite{schafer1996benchmark}. In the cylinder flow case, the boundary condition (BC) describes a cylinder wall in the flow field generally. In \myref{bc of cylinder flow}, the velocity $u$ is limited to zero on the surface of the cylinder wall, where $r$ is the radius of the cylinder wall. The simulation solution reflects the flow in \fref{fig:c_cylinder}.
\begin{equation} \label{bc of cylinder flow}
    u\big(r \cdot cos\theta, r  \cdot sin\theta\big) = (0, 0), \ \theta \in [0, \ 2\pi]
\end{equation}

In recent years, many scholars have employed NN to address both the direct and inverse problems associated with cylinder flow. Though it is an under-determined problem, the velocity field restoration from concentration field is solved effectively by NN \cite{raissi2020hidden}. In the knowledge discovery field, NN can abstract the N-S equation from the velocity field of cylinder flow with high precision \cite{rudy2017data}. To a certain extent, these works show that NN can describe the solution of the N-S equation. However,  current researches heavily relies on the quality and quantity of observation data, thus training based on low quality (i.e., noisy and sparse) data is still an open question.

In general, using physical constraints to improve the NN modeling ability of observation data is a powerful method. However, this approach necessitates high-quality and abundant data. When it comes to the modeling of real measurement data, it becomes crucial to effectively model the sparse and noisy data.

%%% 0214 0014
\subsection{Current challenges}
Obtaining a satisfactory approximate solution solely through PDE constraints can be particularly challenging. The optimization problem of NN is always a non-convex due to the non-linear part inside (optimization of a single hidden-layer NN with non-linear activation function has been proved to be a non-convex problem, \cite{goodfellow2014qualitatively}). Finding the optimal solution with a gradient-based method is NP-hard because the problem is non-convex. As a result, the majority of constructive works typically rely on utilizing more informative data during training (e.g., the observation data of concentration in the full domain on HFM, \cite{raissi2020hidden}).

\begin{figure}
    \centering
    \includegraphics[width=0.7\textwidth]{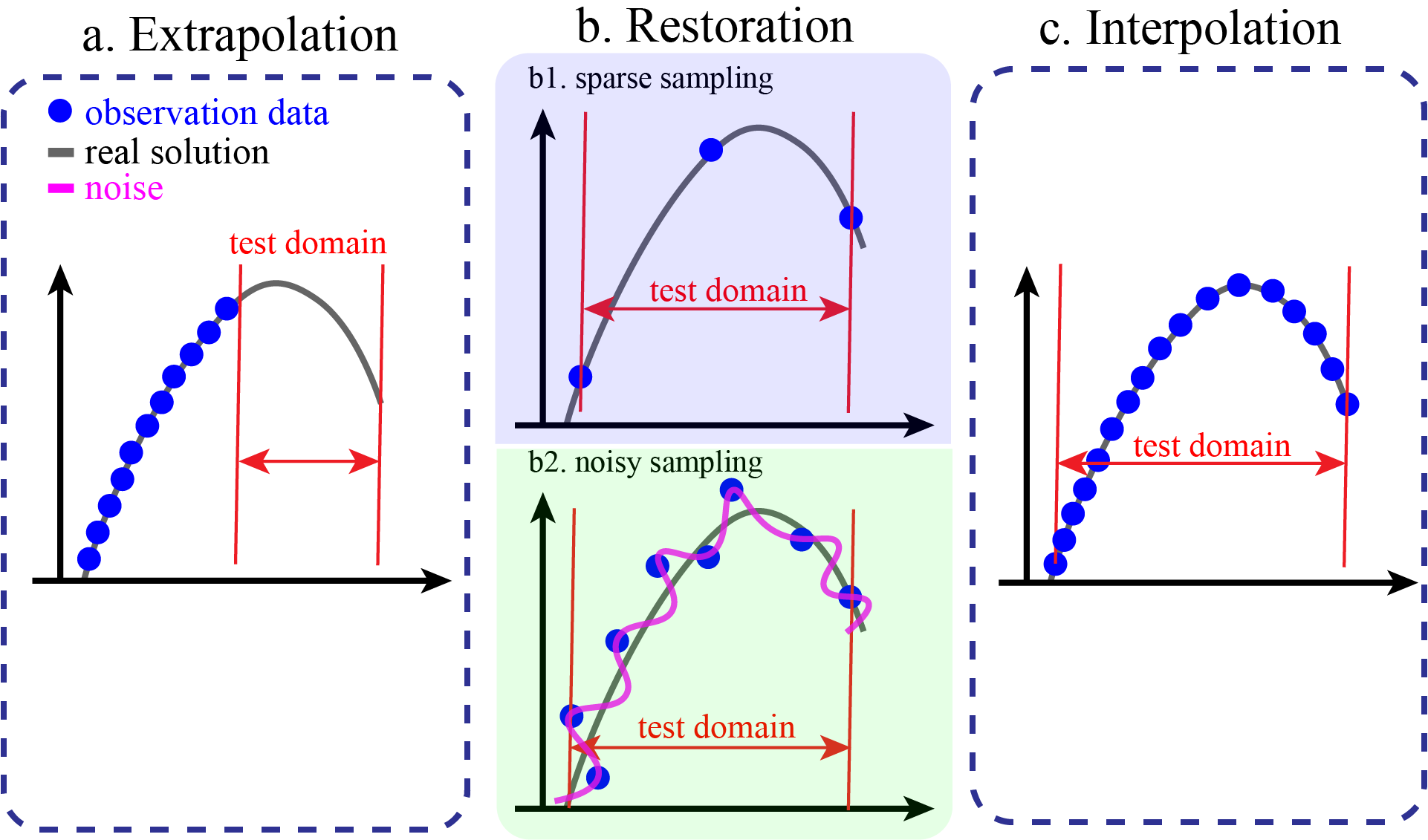}
    \caption{Summary of the physics-informed training tasks. (a) The extrapolation, where known and unknown parts conform to the different data distributions. NN is trained to learn the hidden pattern in the known part, and extrapolate it in the unknown part. (b) The restoration, of which the given and unknown parts conform to the same data distribution. Meanwhile, the data has poor quality and quantity. b1 shows the restoration from sparse data, and b2 is the restoration from noisy data. (c) sufficient and accurate data with the same distribution, which makes the task an ordinary interpolation problem.}
    \label{fig:3_class}
\end{figure}

According to the relationship between the observation and the test domain, the physics informed problem can be divided into three different tasks (as shown in \fref{fig:3_class}). When the observation points all locate outside the desired interval (red test domain in \fref{fig:3_class}), the task of the NN is essentially extrapolation (\fref{fig:3_class}.a). When the observation points are sampled in the same interval of the test domain, the task is called restoration if the observation data is noisy or sparsely sampled from the observation points (see \fref{fig:3_class}.b), or interpolation if the data is abundant and accurate (see \fref{fig:3_class}.c).
% the task is called restoration if the observation data is noisy or sparsely sampled from the observation points (\fref{fig:3_class}.b) or interpolation if the data is sufficient and accurate (\fref{fig:3_class}.c).

In this paper, we focus on the restoration task (i.e., \fref{fig:3_class}.b) since it is more commonly encountered in practice. The NN is designed to learn the embedding distribution of solutions with noisy and sparse data. With the help of physical constraints, NN can provide more reasonable and accurate modeling results. The meaning of restoration is finding a better method to model the noisy and sparse data in the real experiment. The quantity and quality of data required for modeling can be greatly reduced, resulting in lower costs in practice.

\begin{figure}
    \centering
    \includegraphics[width=0.8\textwidth]{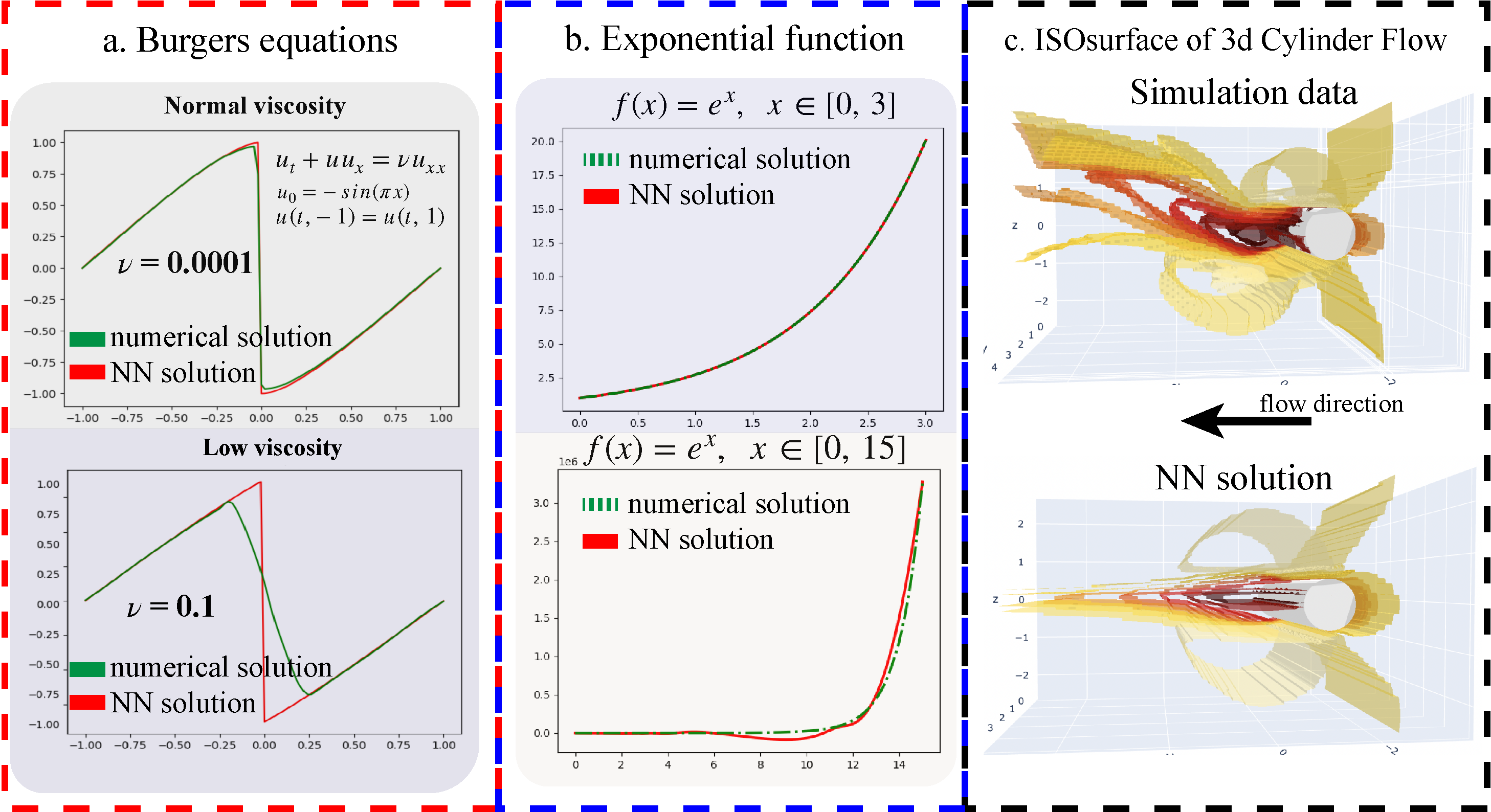}
    \caption{The NN output guided by different equations. (a) The Burgers equation is the classic example in PINN \cite{raissi2019physics}, which is used to show the effect of regularity; (b) The exponential function, which shows the difficulty of NN to handle the interval with a small loss value. NN always focus on the large loss interval; (c) The N-S equation. Facing the complex solution, NN is tend to be trapped in the local minimum and obtain the trivial solution.}
    \label{fig:failed example}
\end{figure}

As the complexity of the problem increases, a greater amount of  observation data is required to describe the distribution of the exact solution. In \fref{fig:failed example}, the first example of Burgers equation (\fref{fig:failed example}.a) shows NN giving the incorrect solution in the interval with the larger differential term. Solutions of different viscosities show the effect of regularity in PDE solutions, which makes NN tend to give smoother outputs. The second example (\fref{fig:failed example}.b), a simple exponential function, directly shows that even though the equation is infinitely differentiable, NN performs poorly when there is a magnitude gap between the scale of the differential terms. The last example is the cylinder flow (\fref{fig:failed example}.c) under the large Reynolds number ($Re$) condition, the complex velocity field makes NN converge to a trivial solution. The frequent alterations in velocity impede the NN's ability to learn the embedding pattern effectively, leading it to merely yield the mean value to attain a local optimum. 

To summarize, the more complex the problem, the greater the amount of data required for training. The sparse and noisy data still contain limited information about the solution's distribution. However, much of the missing information can be inferred by leveraging the PDE as domain knowledge. Although NN can infer most of the missing value, the observation data is still indispensable because it acts as the fixed points in outputs, which anchor the large-scale solutions. The main area for improvement is how to use less or noisier observation data to train a more robusut NN. More specifically, the current challenges of physics-informed training can be itemized as follows:
\begin{itemize}
\item[$\bullet$] The experiment always produces sparse quantities of data when the observation points are gathered through measurements. Lack of data causes NN solutions to frequently yield trivial solutions in complex problems (e.g., \fref{fig:failed example}.c).
\item[$\bullet$] The measurement noise makes observation data out of line with the underlying governing equations. The noise in observation (especially at initial points) can seriously affect the NN optimization process.
\end{itemize}

In order to abstract more information from low-quality and low-quantity data, the equation constraints are improved. In this paper, we propose a new surrogate constraint for the conventional PDE loss using deep learning and numerical method filtering. The surrogate constraint takes the advantage of the mesh-less feature of NN, calculates the PDE loss with the filtered variables instead of the original PDE. The proposed method can be regarded as an intermediary layer based on filter operations, which maintains the equation's form, and is unaffected by the NN architecture. Our proposed filtered PDE model (referred to as FPDE) in this paper shows the following contributions:
\begin{itemize}
\item[$\bullet$] The study proposes a filtered PDE framework that is inspired by large eddy simulation (LES). The solution of physics-informed training is more robust under the constraints of the proposed framework.
\item[$\bullet$] In the noise experiments, the FPDE model can obtain the same quality solution with 100\% higher noise than the baseline.
\item[$\bullet$] In the sparsity experiments, the FPDE model can obtain the same quality solution with only 12\% quantity of observation points of baseline.
\item[$\bullet$] In the real-world experiments with missing equations, the FPDE model can obtain a more physically reasonable solution.
\end{itemize}

In essence, using FPDE as surrogate constraints significantly enhances the ability of neural networks to model data distributions, particularly when dealing with noisy and sparse observations. This improvement is considerable, especially for the work relying on experimental data.

\section{The motivation and ‘conflict’ theory}

To enhance the modeling of noisy and sparse data, a promising avenue for improvement is the co-optimization between PDE and data loss. We discovered that the discrepancy between the directions of the PDE loss and the data loss contributes to some of the challenges in NN optimization. In this paper, the aforementioned challenge is defined as ‘conflict’ and its mathematical derivation is provided and discussed. To mitigate the impact of this conflict, we introduce an improved method inspired by a classic numerical approach (i.e., large eddy simulation, \cite{smagorinsky1963general}). In this part, the incompressible N-S equation is used as an example for demonstration.

\subsection{Mechanism of ‘conflict’} \label{conflict mechanism}

The deep neural network, usually optimized by the gradient-based method, provides a solving paradigm with a large number of parameters. With the classic loss function (i.e., mean squared error, MSE), the solving PDE by NN can be described as a soft-constraint method. Different from the hard-constraint numerical method, the soft-constraint method doesn’t constrain the calculation in each step, but only optimizes the final solution by loss function \cite{chen2021theory}. Under the soft constraint, the optimization process of the parameters in NN is essential. The challenges within the physics-informed framework can be explained from the perspective of optimization.

In the physics-informed framework, the loss function can be written as \myref{general loss}. The IC/BC loss is determined by the initial and boundary conditions, while the PDE loss is computed using the collocation points, which are specific points solely utilized for PDE calculation and remain independent of the observation data distribution. Given the PDE, initial, and boundary conditions, and the observation data jointly determining the final loss, NN training can be broadly defined as the co-optimization of these losses.

%xxx In the co-optimization, the weights of the losses are hyper-parameters. Affected by the dataset, NN architecture and other hyper-parameters, the weights that make the training results better will also be different. Therefore, the tuning of hyper-parameters is also an inevitable part of the PINN research. There are many works that show the importance of weight modification in the training of PINN(e.g., \cite{wang2021understanding} shows the possible solution of tuning the hyper-parameters, and give the introduction and mechanism in \cite{wang2022and}). \cite{cuomo2022scientific} systematically reviewed the development of PINN and also pointed out that the hyper-parameters (e.g., the learning rate and number of iterations) can improve generalization error. But in this work, the focus is the quantitative improvement of conventional PINN by the proposed method. Therefore, the weights in \myref{general loss} are set to 1 uniformly.

During the co-optimization process, the weighting factors assigned to loss terms are crucial hyperparameters. These weights are influenced by factors such as the dataset, neural network architecture, and other hyperparameters, and therefore, the optimal weights may vary accordingly. Consequently, the tuning of hyperparameters becomes necessary. Numerous studies have highlighted the significance of weight tuning in training PINN. \cite{wang2021understanding} present a potential solution for hyperparameter tuning, and the underlying mechanism is described in \cite{wang2022and}. Furthermore, \cite{cuomo2022scientific} provide a comprehensive review of PINN development and emphasize that hyperparameters, such as the learning rate and number of iterations, can enhance generalization performance. In the context of tuning multiple regularization terms, \cite{rong2022lagrangian} propose an optimization method based on Lagrangian dual approaches. However, in this particular work, our primary focus lies on quantitatively improving conventional PINN using the proposed method. Therefore, the weights in \myref{general loss} are uniformly set to 1, which is a commonly adopted setting in several studies (e.g., \cite{raissi2020hidden}, \cite{cai2021physics}, and \cite{goswami2020transfer}).
\begin{equation} \label{general loss}
    \begin{split}
        Loss = L_{data} & + \omega _1 L_{PDE}+ \omega _2 L_{ICBC}\\ 
    \end{split}
\end{equation}
where $L_{data} = \frac{1}{N} ||y-\hat{y}||_2$ and $L_{PDE} = \frac{1}{M} ||residual \ PDE||_2$. The $L_{ICBC}$ is decided on the task specifically. $y$ and $\hat{y}$ represent the NN output and observation respectively. $\omega _1$ and $\omega _2$ represent the weights of PDE loss and ICBC loss.

The optimization direction of the model is the gradient direction of parameters update. It is jointly determined by the sum of each loss direction during each iteration of the multi-objective optimization in \myref{general loss}. The direction of the loss functions in one optimization step can be written as the partial derivative terms in \myref{direction of optimization}, which shows the effect of the loss function on the parameters of NN.
\begin{equation} \label{direction of optimization}
    \theta_t = \theta_{t-1} - \eta(\frac{\partial L_{data}}{\partial \theta_{t-1}} + \frac{\partial L_{pde}}{\partial \theta_{t-1}} + \frac{\partial L_{ICBC}}{\partial \theta_{t-1}})
\end{equation}
where $\theta_t$ represents the NN parameters in iteration t and $\eta$ represents the learning rate (defined as hyperparameter in advance).

Theoretically, there must exist a group of parameters that make PDE and data loss all close to 0. The closed PDEs have embedding distributions of the solutions, but the distribution may have complex pattern, including high-frequency features that pose challenges for neural network (NN) learning. Meanwhile, the observation data also maps the distribution of exact solutions. The physics-informed framework enables NN to learn both the intricate high-frequency patterns from the PDE constraint and the large-scale distribution from observation data. Therefore, having the same embedding distribution under different constraints is the premise of co-optimization.

In the ideal physics-informed framework, PDE and data constraints can help each other out of the local optimum. But in fact, the relationship between these two losses backfired in many practical problems. In the training of physics-informed task, the optimized function is always the residual form of the governing equations, which is the difference between the left and right hand sides of the equation. The 2-dimensional N-S equations are used (2-d case of \myref{NS classic}) as the example to demonstrate the conflict in the co-optimization process. The PDE and data loss can be written in a 2-dimensional case as:
\begin{equation}
    \begin{split}
        L_{PDE} & = \frac{1}{M} \big|\big| \frac{\partial u_i}{\partial t}+u_j \frac{\partial u_i}{\partial x_j}+\frac{1}{\rho}\frac{\partial p}{\partial x_i} -\nu \frac{\partial^2 u_i}{\partial x_j^2} \big|\big|_2 + \frac{1}{M} \big|\big| \frac{\partial u_i}{\partial x_i}\big|\big|_2\\ 
        & = \frac{1}{M} \big|\big| \frac{\partial u}{\partial t}+(u \frac{\partial u}{\partial x} + v \frac{\partial u}{\partial y})+\frac{1}{\rho}\frac{\partial p}{\partial x} -\nu (\frac{\partial^2 u}{\partial x^2} + \frac{\partial^2 u}{\partial y^2}) \big|\big|_2 \\
       & \ + \frac{1}{M} \big|\big| \frac{\partial v}{\partial t}+(u \frac{\partial v}{\partial x} + v \frac{\partial v}{\partial y})+\frac{1}{\rho}\frac{\partial p}{\partial y} -\nu (\frac{\partial^2 v}{\partial x^2} + \frac{\partial^2 v}{\partial y^2}) \big|\big|_2 \\
       & \ + \frac{1}{M} \big|\big| \frac{\partial u}{\partial x} + \frac{\partial v}{\partial y} \big|\big|_2 \\
       L_{data} & = \frac{1}{N} \big|\big| \boldsymbol{y} - \boldsymbol{\hat{y}} \big|\big|_2 \\ 
    \end{split}
\end{equation}
where $\boldsymbol{x} =(t, \ x_i)=(t, x, y)$ and $\boldsymbol{y} =(u_i, \ p)=(u, v, p)$ represent the input and output of NN respectively. $M$ and $N$ are the number of collocation points and observation points.

Since the discrete data also contains the differential information, observation data adds the embedding PDE information to the training process. When the data is sufficient and clean (as in the situation in \fref{fig:3_class}.c), the observation data is consistent with PDE constraints. The PDE can be represented by observation data as \pref{no conflict}.

\begin{proposition} \label{no conflict}
    The no conflict condition. When the number and distribution of observation data are sufficient to describe PDE solutions and the observation data is accurate, discrete data satisfy PDE constraints. The conflict can be represented as the following equation (the difference between residual form of PDE and observation data). In this condition, the conflict is close to 0.
    \begin{equation*}
        PDE(\boldsymbol{y}) =\frac{u_i^{t+1} - u_i^{t}}{\Delta t} + u_j\frac{u_i^{x_j+1} - u_i^{x_j}}{\Delta x_j} + \frac{1}{\rho} \frac{p^{x_i+1} - p^{x_i}}{\Delta x_i} - \nu \frac{u_i^{x_j+1} - 2u_i^{x_j} + u_i^{x_j-1}}{\Delta x_j^2} \approx 0
    \end{equation*}
    where $PDE(\boldsymbol{y})$ is the residual form value of given PDE. The variables in \pref{no conflict} all from the observation data. 
\end{proposition}
%where $PDE(\boldsymbol{y})$ is the residual form value of given PDE. The variables in \pref{no conflict} all from the observation data. 

When the data is noisy (like the restoration in \fref{fig:3_class}.b2), the observation data inherently incorporates noisy information regarding the embedding solution distribution (given in \pref{noisy conflict}). That is, even if the NN can fit the observation data, the PDE constraint cannot be satisfied.

\begin{proposition} \label{noisy conflict}
    The conflict caused by the noise in observation data. When the noisy data is used in the NN training, the conflict between PDE and data constraints occurs. The noisy data can’t represent PDE well, which makes the residual form value of PDE with noisy data larger than that with accurate data. When the NN tries to fit the value, the PDE constraint loss will increase, which causes the difficulties in co-optimization.
    \begin{equation*}
        \begin{split}
            PDE(\boldsymbol{y+\epsilon}) & =\frac{(u_i^{t+1} + \epsilon) - (u_i^{t} + \epsilon)}{\Delta t} + (u_j + \epsilon)\frac{(u_i^{x_j+1} + \epsilon) - (u_i^{x_j} + \epsilon)}{\Delta x_j} \\
            & \ \ + \frac{1}{\rho} \frac{(p^{x_i+1} + \epsilon) - (p^{x_i} + \epsilon)}{\Delta x_i} - \nu \frac{(u_i^{x_j+1} + \epsilon) - 2(u_i^{x_j} + \epsilon) + (u_i^{x_j-1} + \epsilon)}{\Delta x_j^2} \\
            thus \ & \big|\big| PDE(\boldsymbol{y+\epsilon})\big|\big|_2 \geq \big|\big| PDE(\boldsymbol{y})\big|\big|_2 \approx 0 \ , \ \epsilon \sim N(0, \sigma)
        \end{split}
    \end{equation*}
\end{proposition}

When the observation data is sparse and incomplete, the contained information is not enough to aid NN to learn the large-scale distribution of solutions (\pref{sparse conflict}). Under this condition, NN needs to learn the PDE relationship at collocation points without reference. The search space of optimization will expand significantly from the standpoint of the gradient descent process. Although the observation data is error-free and the actual PDE calculation happens at the collocation points, the optimization directions of PDE and data loss (like partial derivative terms in \myref{direction of optimization}) are always in conflict.

\begin{proposition} \label{sparse conflict}
    The conflict caused by the missing of observation data. When the training data is sparse sampled, the distribution information in observation data is not enough. NN acquire to find the missing value by PDE constraint, thus cause the potential conflict in the optimization process. Theoretically, the $ PDE(\boldsymbol{y_{sparse}}) $ can be closed to 0, but finding the correct solution is similar to solving PDE with no observation data.
    \begin{equation*}
        \begin{split}
            PDE(\boldsymbol{y_{sparse}}) & =\frac{u_i^{T+1} - u_i^{T}}{\Delta T} + u_j\frac{u_i^{X_j+1} - u_i^{X_j}}{\Delta X_j} + \frac{1}{\rho} \frac{p^{X_i+1} - p^{X_i}}{\Delta X_i} - \nu \frac{u_i^{X_j+1} - 2u_i^{X_j} + u_i^{X_j-1}}{\Delta X_j^2} \\
            % where \ & \Delta T, \ \Delta X_i \gg \Delta t, \ \Delta x_i
        \end{split}
    \end{equation*}
    where $\Delta T$, $\ \Delta X_i$ $\gg$ $\Delta t$, $\Delta x_i$.
\end{proposition}

The presence of inevitable conflict introduces ambiguity into the optimization direction. The conflict slows the effective gradient descent and eventually results in an insurmountable local optimum.

\subsection{The inspiration from LES}

In order to model the noisy and sparse data better, it is necessary to find a way to overcome the conflict. When calculating the differential terms, the influence of noise or sparsity should be minimized as much as possible. Similar challenges will also be faced in numerical simulation, thus we can refer to existing methods to overcome the challenges. The filters in the numerical method give us inspiration.

In the numerical simulation method, tackling complex equations invariably entails increased computational costs to uphold accuracy. In the computational fluid dynamics (CFD) field, as the Reynolds number increases, the flow tends to be unsteady and disorderly. Since the simulation range is from the domain scale to the smallest dissipation scale, the computational requirement grows at $ Re^3 $ rate \cite{piomelli1999large}. Under the large Reynolds number condition, direct simulation is unaffordable.

In the realm of fluid dynamics, LES stands out for its efficiency in handling turbulent flows \cite{smagorinsky1963general}. The LES methodology achieves this by substituting the small-scale details with an artificially designed subgrid-scale model, which effectively captures the essence of these scales without the need for excessive computational resources. This approach significantly reduces computational costs while preserving the integrity of the solution. Building upon this concept, our study poses a pertinent question: ‘Can a neural network, while learning the Navier-Stokes (N-S) equations, also act as a filter for variables, potentially enabling the derivation of more accurate large-scale solutions from a lesser amount of observational data, irrespective of the specific subgrid model employed?’ Our work explores this question by integrating NN into the framework of LES, aiming to leverage their filtering capabilities to enhance the solution quality with limited data.
% The large eddy simulation (LES, \cite{smagorinsky1963general}) method provides a way to reduce the computational cost as well as guarantee the accuracy. In LES, the grid size is larger than in direct numerical simulation. The small-scale information in LES is replaced and characterized by the artificially designed subgrid-scale model. Conclusively, the LES method greatly reduces the computational cost and maintains the quality of the solution. xxx In this work, regardless of the subgrid model, if NN can filter variables in the process of learning N-S equations, can it also learn more accurate solutions on a large scale from less observation data?
%The LES equation is written as \myref{LES NS}.
%\begin{equation} \label{LES NS}
%    \frac{\partial \overline u_i}{\partial t} + \frac{\partial \overline {u_iu_j}}{\partial x_j}=-\frac{1}{\rho} \frac{\partial \overline p}{\partial x_i} + \nu\frac{\partial^2 \overline u_i}{\partial x_j^2}
%\end{equation}
%where $\overline u_i , \overline u_j , \overline p$ represent the filtered variables in original equation (i.e., \myref{NS classic})

In the problem of solving PDE via NN, the complex solutions also cause bias and insufficiency of observation data which intensify the conflicts in section \ref{conflict mechanism}. Therefore, it is intuitive to find a surrogate constraint to make the NN ignore the small-scale conflict and focus on the large-scale optimization.
While it may seem natural to apply filtering operations directly to the observation data (i.e., the average pooling or the Gaussian filter in the data pre-processing), such operation, like pooling, loses information in the observation data and be hard to implement when the observation data is sparse and randomly distributed. Inspired by LES, we designed a method which filters the NN output in post-processing and rebuilds the governing equations (N-S equations as the example in \myref{FPDE loss}) rather than simply filter the observation data. The new PDE loss defined by the new governing equation acts as a surrogate constraint for the original PDE loss in the training, which is named as ‘filtered partial differential equations’ (FPDE) loss.
\begin{equation} \label{FPDE loss}
    FPDE(\boldsymbol{\overline y})=\frac{\partial \overline u_i}{\partial t} + \frac{\partial \overline {u_iu_j}}{\partial x_j}+\frac{1}{\rho} \frac{\partial \overline p}{\partial x_i} - \nu\frac{\partial^2 \overline u_i}{\partial x_j^2}
\end{equation}

% 0214 1801
\section{Methodology}

To deploy the filter before the PDE calculation, a new intermediate layer is designed to connect the normal NN outputs and the differentiation module. The proposed layer can be regarded as an explicitly defined layer according to the given equation, which connected after the output layer and dedicated to the computation of differential terms. It facilitates the calculation smooth the small-scale oscillations. In addition to the theoretical analysis (section \ref{theoretical improvement}), three experiments are conducted to verify the proposed method. The sparsely sampled simulation data with artificial noise is used to evaluate the improvement quantitatively, followed by testing with the real cell migration and arterial flow data.

\subsection{Deploying filter after NN inference}

The filter operation in physics always implies the constraint of spatial resolution, which can be defined as the convolutional integral in \myref{general conv}. In this integral equation, G is the filter, and $\phi$ is the objective function.
\begin{equation} \label{general conv}
	\overline\phi(x,t)=\int_{-\infty}^{\infty}\phi(y,t)G(x-y)dy
\end{equation}

\begin{figure}
    \centering
    \includegraphics[width=\textwidth]{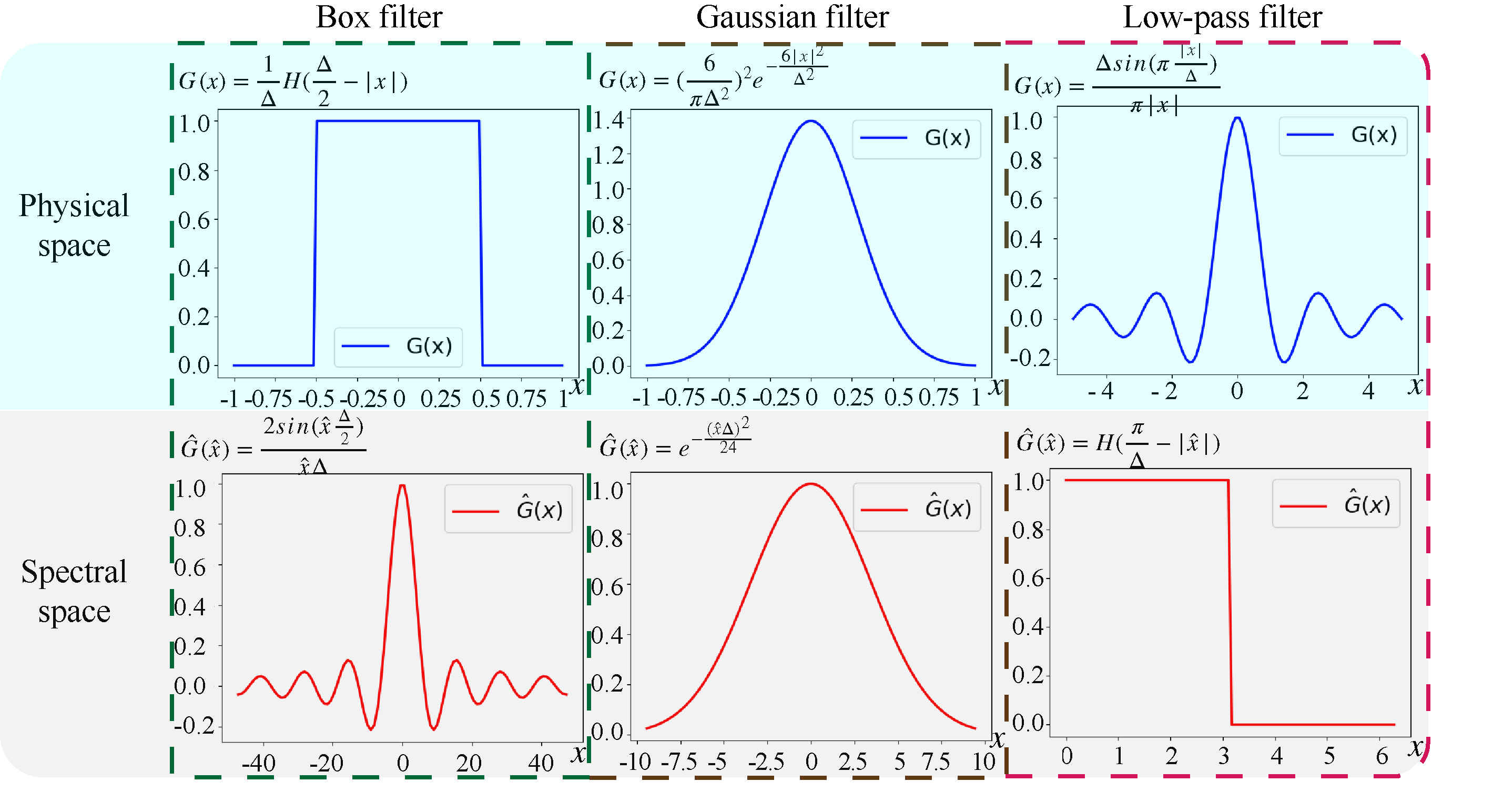}
    \caption{The classic filters in physical and spectral space. The most common Gaussian weight filter kernel will be used in subsequent experiments.}
    \label{fig:filter}
\end{figure}

The classic physical isotropic spatial filters (G in \myref{general conv}) and its Fourier transformation in spectral space is shown in \fref{fig:filter} \cite{cui2004progress}. All filters satisfy the normalization condition to maintain the constants, and the kernel size of filters is represented as $\Delta$. In this paper, all the experiments used the Gaussian kernel. The subsequent experiment verified that the FPDE results are not significantly affected by the type of filter. For details and results of the experiment, please refer to appendix \ref{appendix filter_type}.

\begin{figure}
    \centering
    \includegraphics[width=\textwidth]{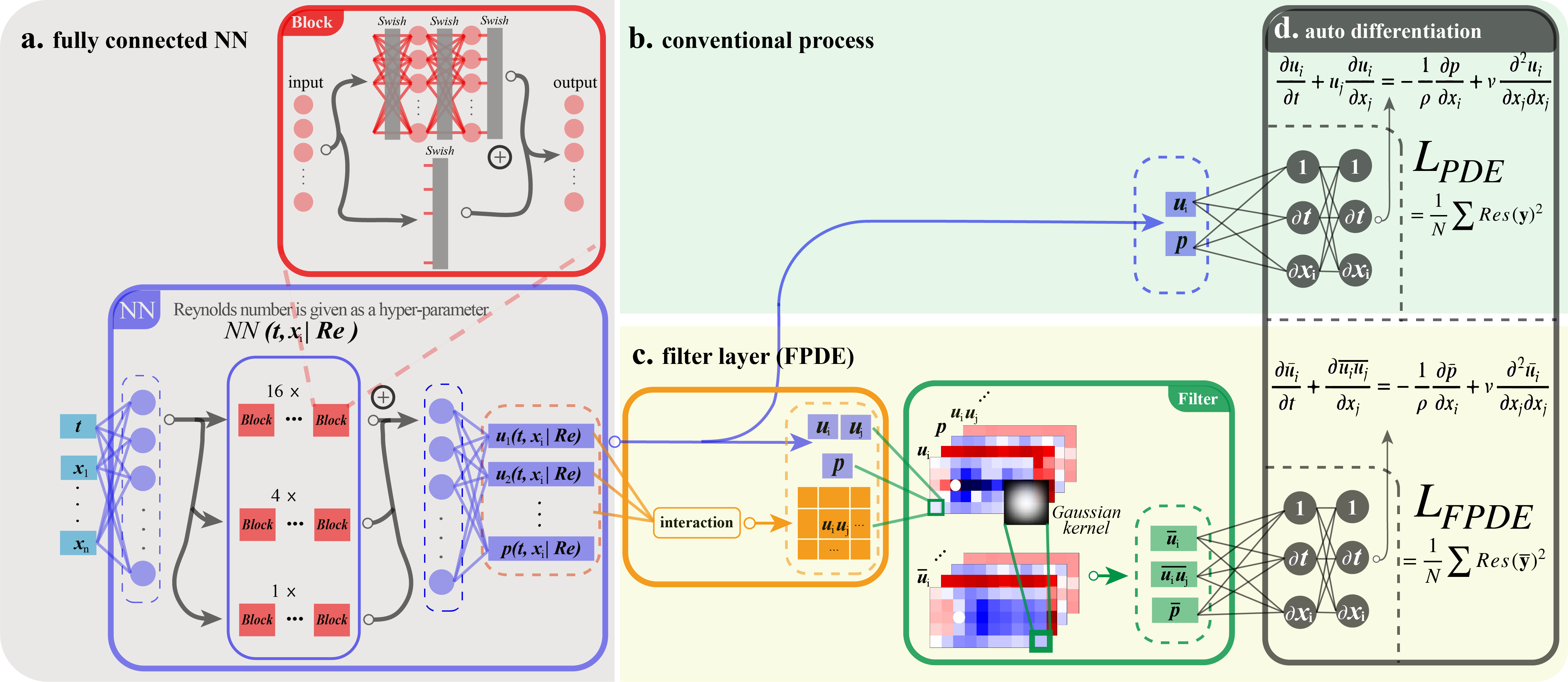}
    \caption{The physics-informed training framework and the proposed FPDE method. The example of solving the N-S equation governed problem. (a) The fully connected neural network. A NN with multiple residual connections is used to model the observation. (b) The conventional PDE loss calculation process, which uses NN outputs in differentiation directly. (c) The improved framework in the PDE loss calculation iteration. The orange part is used to calculate the cross-terms in the equations; the green part is used to filter all variables. The PDEs are calculated with filtered variables instead. (d) The auto differentiation part. Derivatives are automatically generated from the computational graph.}
    \label{fig:model}
\end{figure}

In the context of the computer vision, the filter operation on data is usually called down sampling. When the filter is applied to the PDE calculation, it can be viewed as a new layer in the NN and builds the bridge between the ‘fully connected layer’ and ‘auto differentiation’ (shown in \fref{fig:model}).

The primary  distinction between the regular model and the FPDE model is the calculation of PDE. A multi-layer fully connected network is used as the body. The inputs of NN are the coordinates $x$ and the time $t$. The unknown variables in PDE are the NN outputs (N-S equations as example in \fref{fig:model}). In \fref{fig:model}.b, the conventional process for calculating the PDE loss is depicted, wherein the differential terms are directly obtained from the NN outputs via auto differentiation. The improvements in FPDE are shown in \fref{fig:model}.c: (1) In the orange box, the cross-terms in the N-S equations are calculated before the filter operation. The cross-terms are necessary since the filter operation does not satisfy the associative property of multiplication. Because of the filter operation, the product of two filtered variables is not equal to the filtered result of the product of two variables. (2) The proposed filter, or the defined new activation layer, is shown in the green box. Benefiting from the mesh-free feature of NN, the gridded outputs for the filter operation can be obtained. The $u$ is filtered to $\overline u$ by the Gaussian kernel, and the differential terms are calculated on the filtered outputs. \Fref{fig:model}.d is the calculation of equation constraints. Conventional PINN will use the original outputs to calculate the differential term, while the FPDE will use the filtered outputs. In the calculating of residual form of N-S equations, the Reynolds number of the training data, provided initially, is incorporated into the viscosity term.

In the training of the N-S equations case, a multi-layer NN with residual connections is used. The hidden layer of the NN has three paths with depths of 1, 4, and 16 blocks. The results of these three paths are summed before entering the output layer. Each block consists of 3-layer deep NN with 20 neurons per hidden layer and one residual connection (shown in red box, \fref{fig:model}.a). The activation functions in this NN are all Swish functions.

The details of the training process for FPDE and the conventional PDE model can be summarized in Table.1.
\begin{table}
\centering
\begin{tabular}{lllllllll}
%\toprule
\multicolumn{9}{l}{FPDE and PDE training process (2-d NS equation as an example).}                                                                                                                                                  \\ \midrule
\multicolumn{9}{l}{\begin{tabular}[c]{@{}l@{}}\textbf{Input:} Observation points (x, y, t) →(u, v, p) ; Collocation points ($x_{col}$, $y_{col}$ ,$t_{col}$).\\ \textbf{Constraint:} 2-d N-S equations ; Initial condition (IC) ; Boundary condition (BC).\\ \textbf{Loss:} Mean Squared Error (MSE) loss\end{tabular}}                                                                                                \\
\multicolumn{9}{l}{\textbf{Repeat:}}                                                                                                                                                 \\
        & \multicolumn{8}{l}{\begin{tabular}[c]{@{}l@{}}\textbf{Step 1. Data loss:} for (x, y, t) in observation data\\ \hspace{2em}1.1. Get NN output ($\hat{u},\hat{v},\hat{p}$) by forward propagation.\\ \hspace{2em}1.2. Calculate data loss $MSE((u, v, p), (\hat{u}, \hat{v}, \hat{p}))$, both methods get data loss on the unfiltered field.\\ \textbf{Step 2. FPDE / PDE loss:} for ($x_{col}$, $y_{col}$ ,$t_{col}$) in collocation points\\ \hspace{2em}2.1. Get NN output ($\hat{u}_{col},\hat{v}_{col},\hat{p}_{col}$) by forward propagation.\end{tabular}}                                                                            \\
        & \multicolumn{4}{l}{\hspace{2em} \textbf{FPDE:}}                                                   & \multicolumn{4}{l}{\textbf{PDE:}}                                                \\
        & \multicolumn{4}{l|}{\begin{tabular}[c]{@{}p{20em}@{}}\hspace{2em}2.2. Get the gridded outputs to the 2-d tensor.\\ \hspace{2em}2.3. Filter the 2-d tensor (convolution with Gaussian kernel, i.e., $\overline{u} = \frac{1}{\sum \omega_i} \sum_{i=1}^{n} \omega_i u_i$, $\omega_i$ is the weight in kernel), get the filtered $(\overline u_{col}, \overline v_{col}, \overline p_{col})$.\\ \hspace{2em}2.4. Calculate the differential terms $\frac{\partial \overline u_{col}}{\partial t}$..., etc\\ \hspace{2em}2.5. Calculate FPDE loss of the residual form equations.\end{tabular}} 
        & \multicolumn{4}{l}{\begin{tabular}[c]{@{}p{18em}@{}} 2.2. Calculate the differential terms $\frac{\partial u_{col}}{\partial t}$..., etc\\ \\ 2.3. Calculate PDE loss of the residual form equations.\end{tabular}} \\
        & \multicolumn{8}{l}{\begin{tabular}[c]{@{}l@{}}\textbf{Step 3. IC / BC loss:} for ($u_{ICBC}, v_{ICBC}, p_{ICBC}$) generated by initial / boundary condition\\ \hspace{2em}3.1. Get NN output $ (\hat u_{ICBC}, \hat v_{ICBC}, \hat p_{ICBC} ) $ by forward propagation.\\ \hspace{2em}3.2. Calculate IC/BC loss $MSE((u, v, p) , (\hat u_{ICBC}, \hat v_{ICBC}, \hat p_{ICBC}))$.\\ \textbf{Step 4. Optimization:}\\ \hspace{2em}4.1. Calculate total loss = data loss + equation loss + IC/BC loss\\ \hspace{2em}4.2. Backward propagation and update NN parameters.\end{tabular}}                                                                         \\
\multicolumn{9}{l}{\textbf{Until} the training loss has converged.}                                                                                                                                                  \\ 
%\bottomrule
\end{tabular}
\caption{The FPDE algorithm and comparison with the conventional algorithm.}
\end{table}

The algorithm 1 is the description of \fref{fig:model}. Algorithm 1 compares the difference between the FPDE and classical model more clearly. In both models, the form of the governing equations, initial and boundary conditions, and observation points are completely the same. The sole distinction lies in the FPDE intermediate layer between the forward propagation and auto differentiation parts, as depicted by the orange and green boxes in \fref{fig:model}.b. In order to calculate the filtered PDE, the obtained NN inference outputs need to be gridded. The gridded outputs (i.e., $(\hat u_{col}, \hat v_{col}, \hat p_{col})$, 2-d case) are filtered by the given kernel at step 2.3. The filtered variables (i.e., $(\overline u_{col},  \overline v_{col}, \overline p_{col} )$) are pushed into the auto differentiation part to calculate the differential terms. In the whole FPDE process, the most critical steps are 2.2 and 2.3. Benefit from the mesh-less feature of NN, obtaining the gridded neighbors of given points (i.e., the $NN(x_{col} \pm \Delta x, y_{col} \pm \Delta y, t_{col})$) and acquiring the filtered results is straightforward.

In general, the calculation of FPDE loss can be divided into two steps: calculating cross-terms and filtering. FPDE is transparent for NN architectures and forms of governing equations due to its simplicity. That means FPDE can be applied to most NN architectures and different equations, not just the given example cases.

\subsection{Theoretical improvements of filter} \label{theoretical improvement}
The filtered partial differential equations (FPDE) method is deployed as the surrogate constraint of the original PDE loss. The FPDE constraint helps the optimization process of NN and improves the model's performance in the ‘inference-filtering-optimization’ process. Owing to the intrinsic complexity of NNs, it is challenging to directly demonstrate the specific mechanisms through which FPDE exerts its influence. In this section, we aim to provide potential avenues of explanation and propose a putative mechanism that elucidates the underlying workings of FPDE within the optimization framework of NNs.

\subsubsection{Improvements in problems with the noisy data}
Training with noisy data (like the restoration in \fref{fig:3_class}.b2), the FPDE shows increased accuracy owing to the anti-noise ability of the filter operation. As the basic filter with the Gaussian kernel, the filtered output has a smaller variance than the original output, according to the Chebyshev's inequality \cite{saw1984chebyshev}. To illustrate this, we utilize normally distributed noise, N(0, $\sigma$), which is a commonly used unbiased noise. After the application of the filtering process, the FPDE utilizes data with reduced noise levels compared to the original, pre-filtered data. 
% It is observed that the FPDE loss value is closer to the noise-free value when compared to the original loss value (\myref{FPDE in noisy}).
% To illustrate this, normally distributed noise N(0,$\sigma$), a common unbiased noise, is employed. The FPDE loss value is observed to be closer to the noise-free value compared to the original loss value (\myref{FPDE in noisy}).
\begin{equation} \label{FPDE in noisy}
    \begin{split}
        FPDE(\boldsymbol{\overline{y+\epsilon}}) & =\frac{(\overline{u_i^{t+1} + \epsilon}) - (\overline{u_i^{t} + \epsilon})}{\Delta t} + 
        	\frac{\partial \overline{(u_i + \epsilon)(u_j + \epsilon)}}{\partial x_j} 
%        (\overline{u_j + \epsilon})\frac{(\overline{u_i^{x_j+1} + \epsilon}) - (\overline{u_i^{x_j} + \epsilon})}{\Delta x_j} \\
     	\ + \frac{1}{\rho} \frac{(\overline{p^{x_i+1} + \epsilon}) - (\overline{p^{x_i} + \epsilon})}{\Delta x_i} \\
     	& - \nu \frac{(\overline{u_i^{x_j+1} + \epsilon}) - 2(\overline{u_i^{x_j} + \epsilon}) + (\overline{u_i^{x_j-1} + \epsilon})}{\Delta x_j^2} \\
%        & \ \epsilon \sim N(0,\  \sigma), \ \ \overline{\epsilon} \sim N \left( 0, \  \frac{\sigma \sum\omega_i^2}{\sum^2\omega_i} \right), \ \ 
%        \omega_i = \frac{1}{\sqrt{2\pi}}exp(-\frac{x^2}{2})\\
    \end{split}
\end{equation}
where $\epsilon$ and $\overline{\epsilon}$ represent the noise and filtered noise ($ \epsilon \sim N(0,\  \sigma), \ \ \overline{\epsilon} \sim N \left( 0, \  \frac{\sigma \sum\omega_i^2}{\sum^2\omega_i} \right) $), $\omega_i$ represents the weight in filter ($ \omega_i = \frac{1}{\sqrt{2\pi}}exp(-\frac{x^2}{2}) $). With the Gaussian kernel as the filter operator, the variance of noise decreases at the rate of $n^{-1}$ (n is the size of given Gaussian kernel). This indicates that just a small kernel can greatly reduce the interference of noise. 

\begin{figure}
    \centering
    \includegraphics[width=\textwidth]{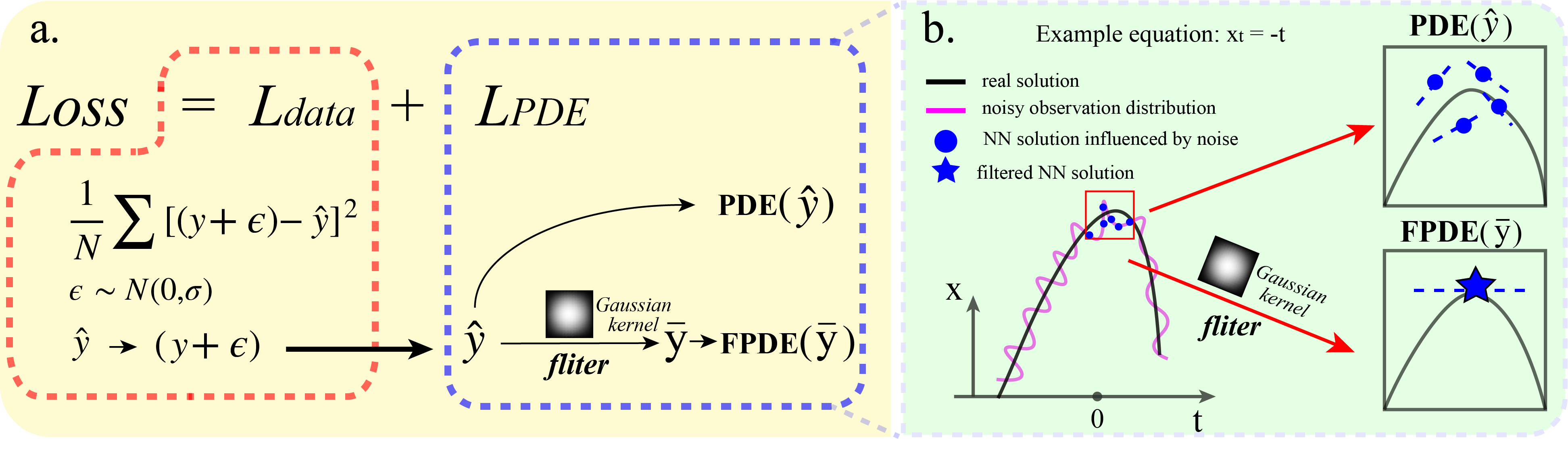}
    \caption{Given possible mechanism of the FPDE method for processing noisy data. (a) Though the noisy data doesn't directly participate in the PDE loss calculation, it still affects the PDE loss by changing the NN output from the previous iteration. (b) The filtered NN outputs reduce the noise, making PDE calculations more robust and accurate.}
    \label{fig:noisy improvement}
\end{figure}

Observational data contaminated with noise can significantly diminish the precision of predictions generated by neural networks (NNs). The effects and directional trends of such noise are illustrated in \fref{fig:noisy improvement}.a, where it is evident that data degradation progressively aligns the NN's output with the noisy data set during each iterative cycle. This compromised output, subsequently influenced by the noise, can introduce bias into the computation of PDEs.

\Fref{fig:noisy improvement}.b presents a straightforward example, demonstrating that the output refined through a filtering process (represented by the blue star) more closely approximates the actual solution. This refined output point yields an accurate derivative, which is instrumental in mitigating the NN's susceptibility to noise. In contrast, outputs from a standard PDE model (depicted as blue dots) are subject to the distorting effects of noise (represented by the pink line), thereby failing to furnish the correct derivatives necessary for precise calculations.

In summary, the implementation of a filtering mechanism effectively diminishes the magnitude of noise interference. As a result, FPDE are capable of yielding superior modeling outcomes even when the input data are noisy. It is hypothesized that the enhanced accuracy in the computation of differential terms is attributable to the reduced presence of noise within the filtered data set.
% Noisy observation data can lead NN to produce inaccurate predictions. The influence and tendency of noise are shown in \fref{fig:noisy improvement}.a, the data loss leads the NN output close to the noisy observation data in each iteration. Then, the NN output, which is influenced by the noisy data, will bias the calculation of PDE. In \fref{fig:noisy improvement}.b, a simple example shows that the filtered output (blue star) is closer to the real solution, the filtered star point provides the correct derivative to help NN resist the noise. In a conventional PDE model, the outputs (blue dots) are influenced by the noise (pink line), and can not provide the correct derivatives. In conclusion, the filtering operation reduces the value of noise. FPDE can have better modeling results under noisy data. The reason is speculated that there is small noise in the filtered data, which makes the calculation of differential terms more accurate.

\subsubsection{Improvements in problems with the sparse data}
In the context of sparse data problems, the enhancement brought by FPDE can be viewed as an inverse application of the aforementioned mechanism. Within \myref{FPDE in noisy}, the discrepancy in the optimization trajectory between the dataset and the PDE loss is mitigated through a filtering process. This intervention allows the NN to function without being impeded by the interplay of these two forms of loss. In section \ref{conflict mechanism}, it is hypothesized that the source of discord stems from the suboptimal quality of the input data. Empirical findings indicate that the FPDE approach persists in its optimization endeavors despite heightened conflict when training with sparse data. From a co-optimization perspective, the attenuation of this conflict can be interpreted as an inverse decoupling of the data and PDE loss elements.
% The improvement of training with sparse data can be supposed as the inverse application of the possible mechanism above. In \myref{FPDE in noisy}, the divergence in the optimization direction between data and PDE loss is reduced by filtering, enabling the NN to operate without being constrained by the interdependence of these two losses. In section \ref{conflict mechanism}, we suppose that the conflict arises from low-quality data. When using sparse data for training, the FPDE method was found to continue to optimize under greater conflict. In the view of co-optimization, the reduction of conflict can be regarded as the decoupling of the data and PDE loss inversely.

In comparing the FPDE model with the conventional PDE model, particularly with respect to sparse data scenarios, the PDE loss takes a straightforward residual form that directly constrains the solution. Conversely, the FPDE model operates on the filtered ‘mean’ solution, as detailed in \myref{FPDE sparse improvement}. Relative to the original problem, the FPDE-governed problem introduces additional variables that can be autonomously learned. When computing the equation's residual, the FPDE approach involves a greater number of NN inference outputs. Essentially, this allows for a broader spectrum of NN output values corresponding to the same filtered outcome, suggesting a richer set of potential solutions. From this vantage point, the FPDE effectively moderates the coupling between the PDE loss and data loss throughout the training process.
% xxx In the comparison of the FPDE model and the conventional PDE model with sparse data, the simple residual form of the PDE loss constrains the solution directly, while the FPDE constrains the filtered ‘mean’ solution (\myref{FPDE sparse improvement}). Compared with the original problem, the FPDE-governed problem has more variables that can be learned aotonomously. When calculating the residual of equation, FPDE will calculate it through more NN inference outputs. In other words, the neural network outputs can have more types of value corresponding to the same filtered result, indicating a greater variety of possibilities. From this perspective, it can be speculated that FPDE partially decouples the PDE loss and data loss during the training process.
\begin{equation} \label{FPDE sparse improvement}
    \begin{split}
        FPDE(\boldsymbol{\overline{y_{sparse}}}) & =\frac{\overline{u_i^{T+1}} - \overline{u_i^{T}}}{\Delta T} + \frac{\partial \overline{u_i u_j}}{\partial X_j} + \frac{1}{\rho} \frac{\overline{p^{X_i+1}} - \overline{p^{X_i}}}{\Delta X_i} - \nu \frac{\overline{u_i^{X_j+1}} - 2\overline{u_i^{X_j}} + \overline{u_i^{X_j-1}}}{\Delta X_j^2} \\
        & \overline{u} = \frac{1}{\sum \omega_i} \sum_{i=1}^{n} \omega_i u_i, \ \omega_i = \frac{1}{\sqrt{2\pi}}exp(-\frac{x^2}{2}) \\
%        \ for \ & PDE(\boldsymbol{y_{sparse}}), \ df_{sparse}=k - m = 0 \\
%        \ for \ & FPDE(\boldsymbol{\overline{y_{sparse}}}), \ df_{\overline{sparse}}=n * k - m \geq 0
    \end{split}
\end{equation}
where $n$ is the filter size, the filtered variables are calculated from multiple NN outputs.

\subsection{Design of experiments}

Three experiments—the cylinder flow, cell migration, and artery flow—are used to demonstrate the performance of FPDE on sparse and noisy data. In the cylinder flow case, the improvements of FPDE with simulation data are verified quantitatively; in the cell migration case, we evaluate the FPDE's ability to correct real data when equations have missing coefficients; in the arterial flow case, we assess the performance of FPDE with inconsistent equations and observation data. 

\subsubsection{Simulation data of cylinder flow}

To verify the improvement of FPDE under sparse and noisy training data, the sparse dataset and noise dataset are designed for the quantitative experiment. In the experiment, the sampling ratio and noise level can be controlled to quantitatively demonstrate the improvement effect of FPDE.

We designed two experiments to verify the FPDE improvements in sparsity and noise of training data. In the generation of sparse data in Group 1, the datasets are randomly sampled by the decreasing sampling ratios in Table.1. To demonstrate the NN restoration ability under various levels of data missing, seven datasets of different sizes are employed in Group 1. Obviously, the less observation it has, the more inaccuracy it produces.In Group 2, noise is added to the $[u, v, p]$ field to make training more difficult (as demonstrated in \pref{noisy conflict}). The ‘additive white Gaussian noise’ (AWGN), the most common noise in the noise analysis, is chosen as the artificial noise added in $2^{-10}$ sampled dataset. The variances of the noise in $[u, v, p]$ are jointly decided by the standard deviation in $[u, v, p]$ and the noisy rate r ($\epsilon_u = N(0, r \cdot std_u)$). Seven datasets of different noisy levels are used as the variables of Group 2 to show the flow restoration ability with the different data error levels. As anticipated, higher levels of noise result in increased inaccuracies in the restoration process.

\begin{figure}
    \centering
    \includegraphics[width=0.5\textwidth]{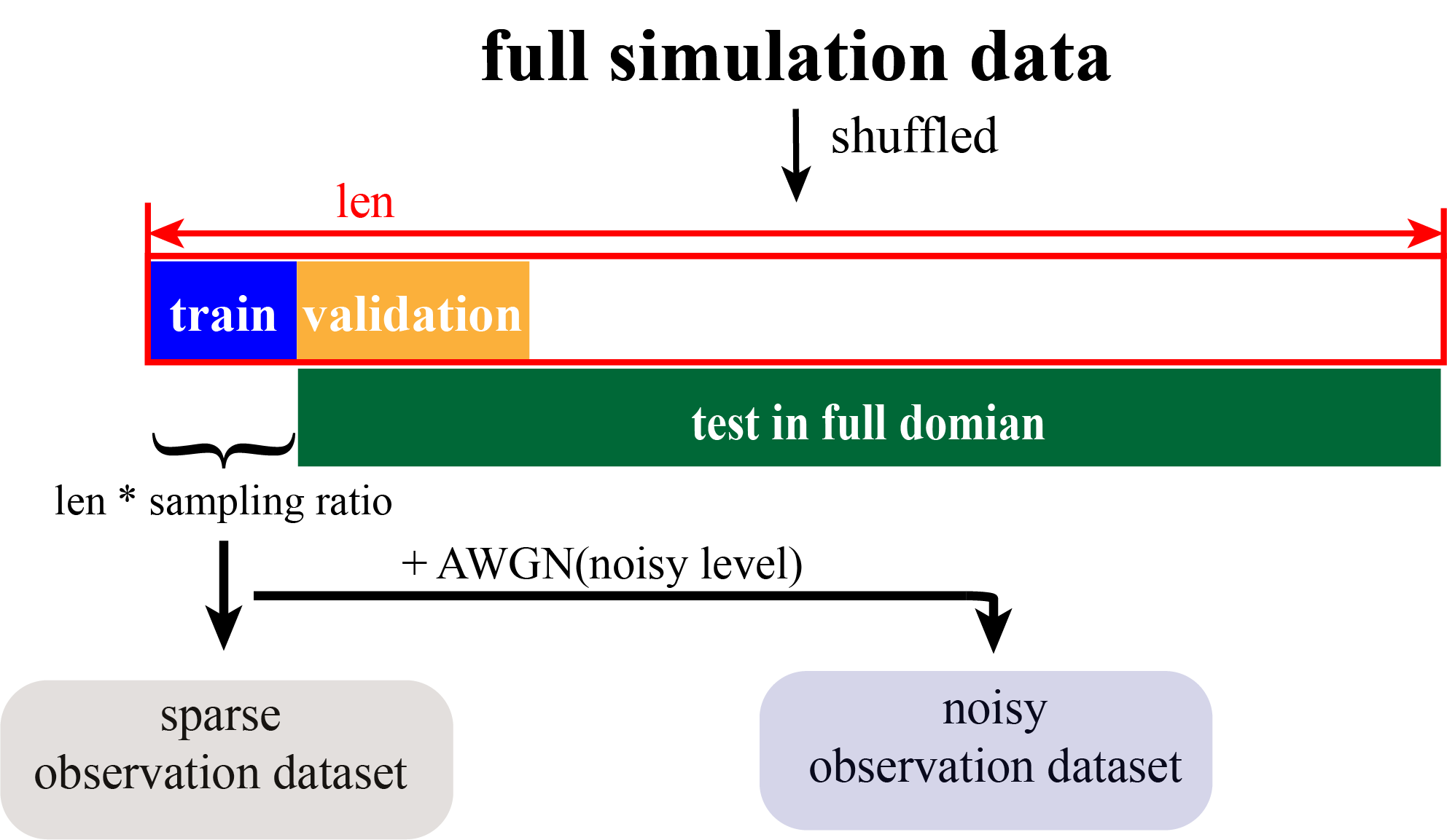}
    \caption{The pre-possessing in full simulation data. In the experiments, the training data is randomly sampled in full dataset. In sparse group, the sampling ratio is the adjustable variable. In noisy group, the noisy level is the adjustable variable.}
    \label{fig:separation}
\end{figure}

The entire simulation data is divided into 3 parts: the training, validation, and test datasets. To test the restoration ability, the restoration outputs are plotted across the entire domain. The division is shown in \fref{fig:separation}.

\begin{figure}
    \centering
    \includegraphics[width=\textwidth]{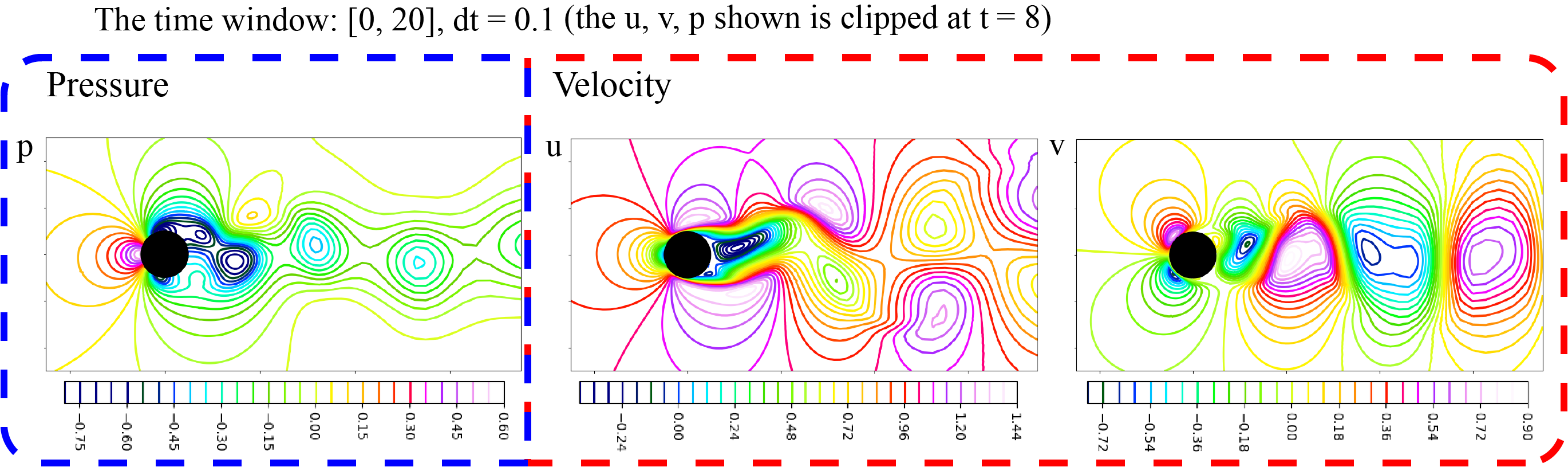}
    \caption{The numerical simulation data of the cylinder flow problem. The training data is sampled from the original data and contaminated with noise. Similar to the boundary condition, the black circle is the cylinder wall.}
    \label{fig:iso surface}
\end{figure}

For details of cylinder flow, data sources, and pre-processing methods, please refer to appendix \ref{appendix a}. \Fref{fig:iso surface} is an overview of the simulation data.

Finally, experiments are conducted in two groups to evaluate the effects of sparsity and noise. The details of the two groups of datasets are presented in Table 2. For each experiment, the FPDE and the baseline model are trained in parallel to show improvements. The processes of data sampling and adding noise are also described in appendix \ref{appendix a}.

\begin{table}
\centering
\begin{tabular}{m{9em}m{4em}m{4em}m{4em}m{4em}m{4em}m{4em}}
%\toprule
\textbf{Base data} & \multicolumn{6}{c}{\textbf{Group and Data overview}}                                                                                        \\ \midrule
          & \multicolumn{6}{c}{Group 1: sparse data sampled by different sampling ratio}                                                       \\
full size & $2^{-8}$                    & $2^{-10}$                    & $2^{-12}$                    & $2^{-13}$                   & $2^{-14}$                  & $2^{-16}$                  \\
          & \multicolumn{6}{c}{\begin{tabular}[c]{@{}c@{}}Group 2: noisy data generated by different noisy\\  standard deviation\end{tabular}} \\
$2^{-10}$ sampled data       & $25\% \cdot std$                    & $50\% \cdot std$                    & $75\% \cdot std$                    & $100\% \cdot std$                   & $125\% \cdot std$                  & $150\% \cdot std$                  \\ 
%\bottomrule
\end{tabular}
\caption{Summary of two groups of data. These groups of data are used to train the FPDE and conventional PDE models and validate their flow restoration abilities under sparse and noisy conditions.}
\end{table}

The evaluation criteria are defined in \myref{cf loss}. The conventional PDE model and FPDE model are trained and tested on the same dataset. 
\begin{equation} \label{cf loss}
	\begin{split}
		Loss & = \frac{1}{N}\big|\big| \boldsymbol{y} - \boldsymbol{\hat{y}} \big|\big|_2  + \frac{1}{M}\big( \sum_{i=1}^{3} \big|\big|  e_i \big|\big|_2 \big)\  + \frac{1}{W} \big|\big| \frac{\partial^k \boldsymbol{y}_{IC/BC}}{\partial \boldsymbol{x}^k_{IC/BC}} - \frac{\partial^k \boldsymbol{\hat y}_{IC/BC}}{\partial \boldsymbol{x}^k_{IC/BC}} \big|\big| \\
        e_1 & = \frac{\partial u}{\partial t}+(u \frac{\partial u}{\partial x} + v \frac{\partial u}{\partial y})+\frac{1}{\rho}\frac{\partial p}{\partial x} -\nu (\frac{\partial^2 u}{\partial x^2} + \frac{\partial^2 u}{\partial y^2}) \\
        e_2 & = \frac{\partial v}{\partial t}+(u \frac{\partial v}{\partial x} + v \frac{\partial v}{\partial y})+\frac{1}{\rho}\frac{\partial p}{\partial y} -\nu (\frac{\partial^2 v}{\partial x^2} + \frac{\partial^2 v}{\partial y^2}) \\
        e_3 & = \frac{\partial u}{\partial x} + \frac{\partial v}{\partial y} 
	\end{split}
\end{equation}
where $N,M,W$ represent the numbers of observation, collocation and IC/BC points in one iteration, respectively. $e_1,e_2,e_3$ represent the values of residual form PDE (N-S equations as example). $\frac{\partial^k}{\partial \boldsymbol{x}^k_{IC/BC}}$ represents the paradigm of boundary conditions in different task (e.g., k=0 / 1 means the Dirichlet / Neumann boundary condition).

\subsubsection{Measurement data of cell migration}

In this experiment, the real-world measurement data is used to demonstrate the improvement of FPDE in real-world situations. Generally, there are two difficulties when using real data in this experiment. The first challenge stems from the high noise in the observation data. In the experiments, the measurement data is mainly obtained by sensors or manual measurements, which means the data is always noisy and sparse. When the measurement data is used as observation points in physics-informed framework, it leads to conflict between data distribution and theoretical equation. The cell number $C$ has high noise because the experimental data is automatically collected by the CV algorithm. The second challenge in this experiment is the missing coefficients in the equations. Since some coefficients of the equation are unknown, NN predicts those segments without collocation points. 

The cell migration data in reproducibility of scratch assays is affected by the initial cell density in the given scratch \cite{jin2016reproducibility}. It shows the relationship of cell distribution in scratch assays with time, space, and initial cell density. The data elucidates that when a scratch occurs, cells migrate to repair the scratch. Existing theories often use the Fisher-Kolmogorov model to describe the process of collective cell spreading, expressed as:
\begin{equation} \label{cm equation}
	\frac{\partial C}{\partial t} = D\frac{\partial^2 C}{\partial x^2} + \lambda C \big[ 1 - \frac{C}{K} \big]
\end{equation}
where the dependent variable $C$ represents the cell concentration. $K, \lambda \ and \ D$ represent the carrying capacity density, the cell diffusivity and the cell proliferation rate respectively. In this context, $K, \lambda \ and \ D$ can be viewed as the coefficients decided by initial cell density (n). Because this is a variable coefficient equation, it cannot calculate the unknown PDE at any collocation points. In the experiment described in this paper, the coefficients in $n=14,000$ and $n=20,000$ are known. The aim is to model the $n \in (14,000, 20,000)$ interval data through the FPDE training.

\begin{figure}
    \centering
    \includegraphics[width=0.8\textwidth]{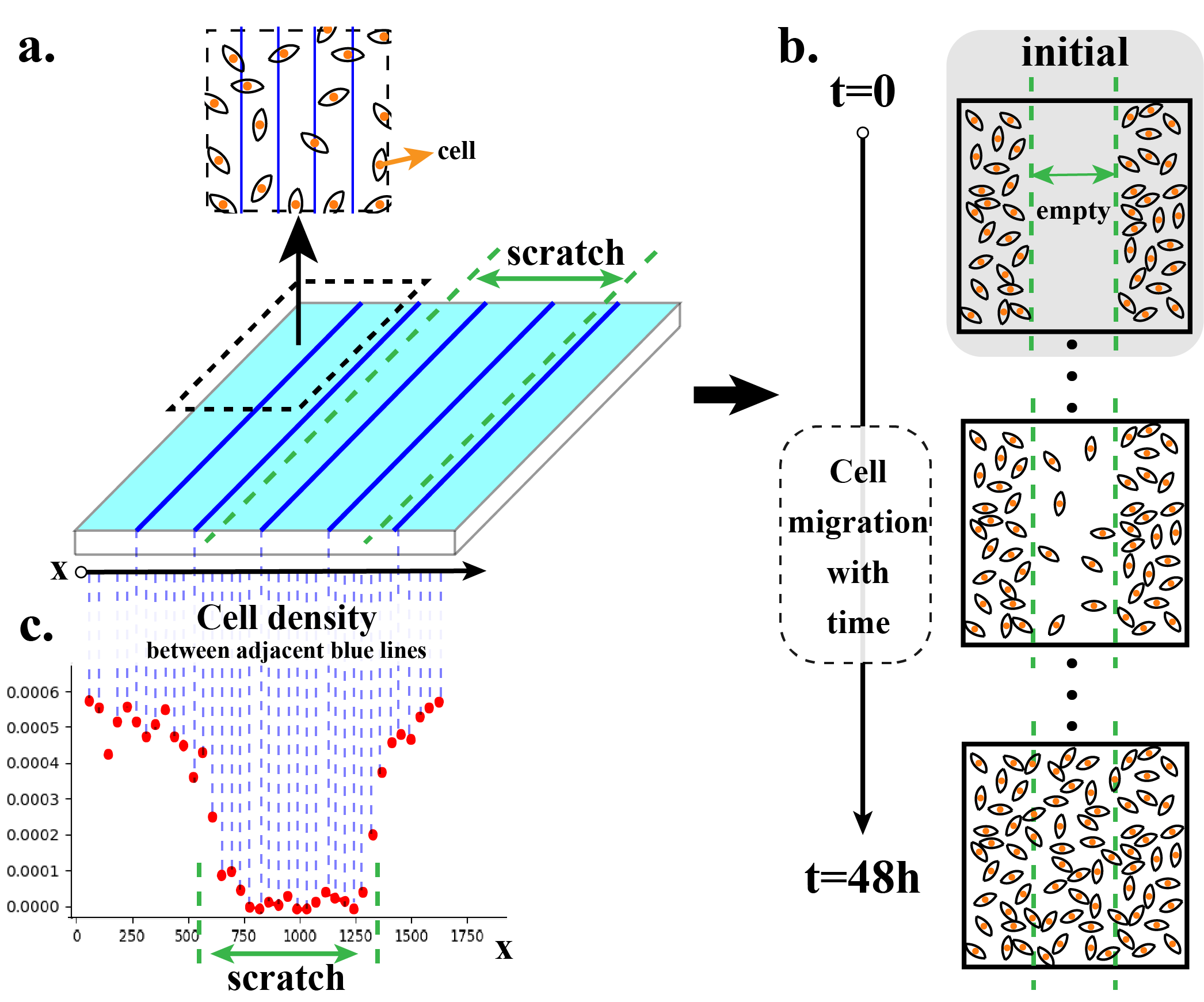}
    \caption{The experiment of cell migration and data collection. (a) The square petri and counting method of blue lines. (b) Cells fill in the scratch by migrating. (c) Measurement data distribution.}
    \label{fig:cm}
\end{figure}

The \fref{fig:cm} below is a schematic diagram of the cell migration experiment. For more details on the experiment, coefficients, and dataset distribution, please refer to appendix \ref{appendix b}.

In summary, the NN models the mapping relationship ‘$NN(t, x, n) \rightarrow C$’. And the final loss function in \myref{general loss} can be written as follows (\myref{cm loss}). When it comes to the calculation of FPDE, the variable $C$ is filtered first and calculated in the same form as \myref{cm loss}. Both the conventional PDE model and the FPDE model are trained until converged and tested on the same dataset.
\begin{equation} \label{cm loss}
	\begin{split}
		Loss & = \frac{1}{N}\big|\big| \boldsymbol{C} - \boldsymbol{\hat{C}} \big|\big|_2  + \frac{1}{M}\big(\big|\big|  e \big|\big|_2 \big)\  + \frac{1}{W} \big|\big| \frac{\partial^k \boldsymbol{y}_{IC/BC}}{\partial \boldsymbol{x}^k_{IC/BC}} - \frac{\partial^k \boldsymbol{\hat y}_{IC/BC}}{\partial \boldsymbol{x}^k_{IC/BC}} \big|\big| \\
        e & = \frac{\partial C}{\partial t} - 530.39\frac{\partial^2 C}{\partial x^2} - 0.066C + 46.42C^2, \ if \ n=14,000\\
        e & = \frac{\partial C}{\partial t} - 982.26\frac{\partial^2 C}{\partial x^2} - 0.078C + 47.65C^2, \ if \ n=20,000\\
        e & = 0, \ else
	\end{split}
\end{equation}
where $N,M,W$ represent the numbers of observation, collocation and IC/BC points in one iteration, respectively. $e$ is the residual form value of \myref{cms}. Because of the changing coefficients, $e$ should be calculated according to three categories ($n=14,000 / 20,000/else$). The constants are obtained by regression in the experiment on \cite{chen2021physics}.

\subsubsection{Measurement data of arterial flow}

When the equation is obtained through the ideal model, there are always significant disparities between the actual situation and the description of the equation. In this experiment, arterial blood flow measurements are used to compare the modeling results of FPDE and baseline with noisy data.

\begin{figure}
    \centering
    \includegraphics[width=1\textwidth]{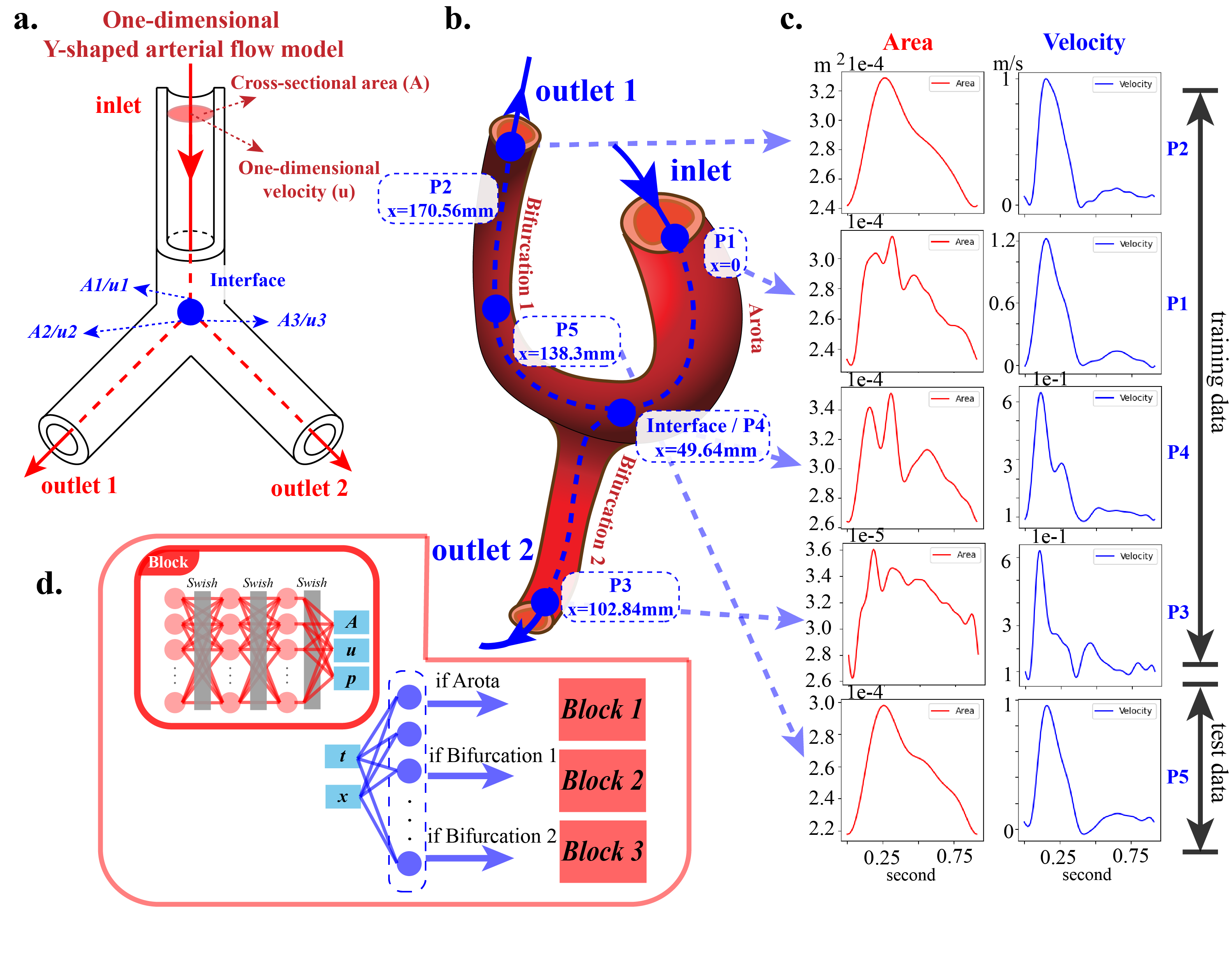}
    \caption{The experiment of one-dimensional blood flow in the Y-shaped artery. (a) The theoretical model (including the definitions of $A, \ A_i$ and $interface$). The \myref{af} used in the training are derived from this model. (b) The location of five measurement points and the schematic diagram of artery. (c) Overview of measurement data (area and velocity). (d) Multi-head NN to predict the velocity in different vessels.}
    \label{fig:af}
\end{figure}

The data regarding arterial flow shows the velocity of blood when it flows through the arterial bifurcation  \cite{kissas2020machine}. The theoretical equations of velocity are shown in \myref{af}. $A$ is the cross-sectional area of the vessel, and $u$ is the axial velocity. Similar to the N-S equations (\myref{NS classic}, $\rho$ and $p$ represent density and pressure, respectively), $A_1$ and $u_1$ are the cross-sectional area and velocity of the interface in the aorta. $A_2, u_2$ and $A_3, u_3$ are the area and velocity of the interface in two bifurcations. The physical relationship between the aorta and two bifurcations is shown in \fref{fig:af}.a, the real-world vessel's shape is shown in \fref{fig:af}.b.
\begin{equation} \label{af}
	\begin{split}
		\frac{\partial A}{\partial t} & = - \frac{\partial Au}{\partial x} , \ \frac{\partial u}{\partial t} + u\frac{\partial u}{\partial x} = -\frac{1}{\rho}\frac{\partial p}{\partial x} \\
		A_1 u_1 & = A_2 u_2 + A_3 u_3 , \ p_1 + \frac{\rho}{2}u_1^2 = p_2 + \frac{\rho}{2}u_2^2 = p_3 + \frac{\rho}{2}u_3^2 \\
	\end{split}
\end{equation}

An overview of the experiment is depicted in the \fref{fig:af}. Notably, a multi-head neural network is used to fit different segments of vessels. Briefly, the aim is to train the NN with measurement data from only four observation points and model the entire blood vessel.
 
\Fref{fig:af} illustrates the experiment of one-dimensional blood flow in the Y-shaped artery. Previous studies modeled the velocity at the bifurcation based on an idealized Y-shaped one-dimensional vessel (shown in \fref{fig:af}.a). The actual blood is not ideal, thus the measurement data always can't fit the embedding distribution in \myref{af} well. The conflict raised in section \ref{conflict mechanism} affects optimization a lot. In the real experiment measuring blood flow, the schematic diagram of an artery is shown \fref{fig:af}.b, with blood directions indicated by blue dotted lines. The data are measured at the five points, which are shown in \fref{fig:af}.b, and the measured variables ($A, u$) are shown in \fref{fig:af}.c. The in-vivo data is measured by the Magnetic Resonance Imaging (MRI) method in Machine learning in cardiovascular flows modeling \cite{kissas2020machine}. All data (area and velocity) are measured within 850ms. Since the data is measured in an open vessel, the boundary condition is unknown in training.
 
In order to model the distribution ‘$NN(t, x) \rightarrow (A, u, p)$’ in three parts of vessel, we build a multi-head NN to predict the $(A, u, p)$ in different parts separately (shown in \fref{fig:af}.d). NN is trained by the data from three endpoints (points 1, 2, and 3) and the interface (point 4). The collocation points are sampled among the entire vessel as constraints. The data in point 5 is reserved for testing. During parallel training, baseline PDE is calculated directly. When calculating FPDE, the variables $y = (A, u, p)$ are filtered first and calculated in the same form as in \myref{af loss}.

The final loss function in \myref{general loss} can be written as follows. Conventional PDE model and FPDE model are trained until converged and tested in same dataset.
 \begin{equation} \label{af loss}
	\begin{split}
		Loss & = \frac{1}{N}\big|\big| \boldsymbol{y} - \boldsymbol{\hat{y}} \big|\big|_2  + \frac{1}{M}\big( \sum_{i=1}^{2} \big|\big|  e_i \big|\big|_2 \big) + \frac{1}{W}\big( \sum_{i=1}^{3} \big|\big|  f_i \big|\big|_2 \big)\ \\
        e_1 & = \frac{\partial A}{\partial t} + \frac{\partial Au}{\partial x}, \ 
        e_2 = \frac{\partial u}{\partial t} + u\frac{\partial u}{\partial x} + \frac{1}{\rho}\frac{\partial p}{\partial x} \\
        f_1 & = A_1 u_1 - A_2 u_2 - A_3 u_3 \\
        f_2 & = p_1 + \frac{\rho}{2}u_1^2 - p_2 - \frac{\rho}{2}u_2^2, \ 
        f_3 = p_1 + \frac{\rho}{2}u_1^2 - p_3 - \frac{\rho}{2}u_3^2 \\
	\end{split}
\end{equation}
where $N,M,W$ represent the numbers of observation, collocation and interface points in one iteration, respectively. $e_1,e_2$ represent the residual value of governing equations, $f_1,f_2$ represent the residual value of the interface constraints which constrain mass and energy conservation in interface.

\section{Results}

To ascertain the superiority of the FPDE constraint over conventional PDE constraint, the performance of the corresponding models are compared in this section via multiple experiments. For comparison, the NN output, the residual map, and the converged losses are plotted. The experimental findings offer evidence of the conflict—aligning with the hypotheses outlined in section \ref{conflict mechanism})—and corroborate that FPDE exhibits superior conflict resistance capabilities, a potential mechanism of which has been introduced in section \ref{theoretical improvement}).
% xxx The experiment results provide evidence of conflict (hypotheses presented in section \ref{conflict mechanism}) and FPDE has better conflict resistance ability (possible mechanism given in section \ref{theoretical improvement}).

\subsection{The comparison between FPDE and baseline model}

\subsubsection{Quantitative analysis of simulation data}

\begin{figure}
    \centering
    \includegraphics[width=\textwidth]{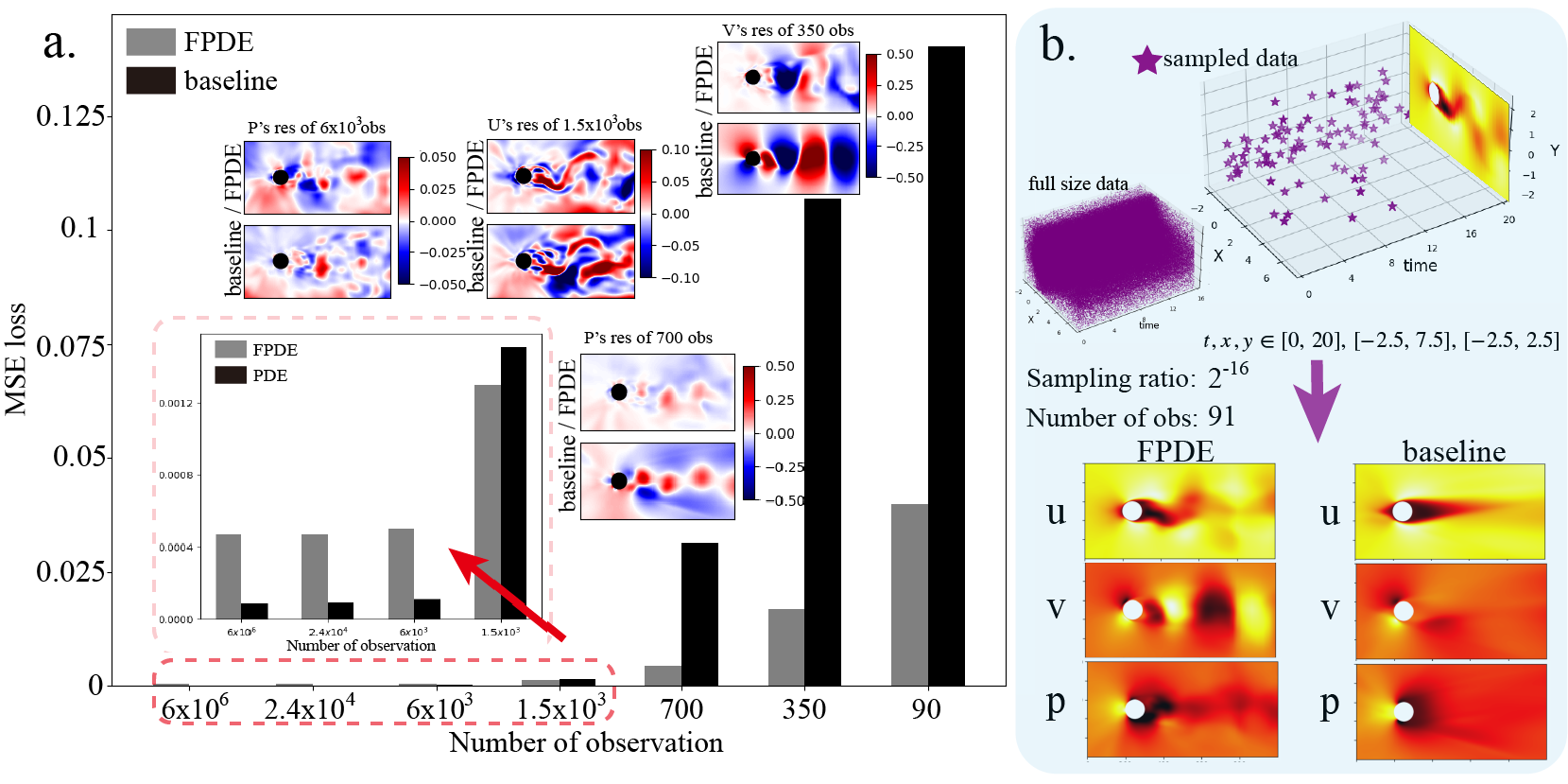}
    \caption{The NN outputs and Mean Squared Error of the FPDE and baseline models. (a) The Mean Squared Error and the residual between the ground truth and output. This result shows the FPDE method works better with sparse observation. (b) A typical example with a $2^{-16}$ sampling ratio. The comparisons of NN outputs ($u,v,p$) are below.}
    \label{fig:sparse result}
\end{figure}

In the sparse data experiment, 7 pairs of converged losses are shown in the histogram of \fref{fig:sparse result}.a. The simulation data is used as the ground truth in this experiment. The x-axis of this histogram is the number of simulation data used in the training dataset, and the bar on the y-axis is the MSE on test data. The residuals between the ground truth and NN restored flow field are plotted above the loss bar, where the top is the FPDE and below is the baseline residual. According to the color map in the residual plot, a whiter color means a solution that is closer to the ground truth. In \fref{fig:sparse result}.a, the residuals of FPDE are less than baseline generally. This improvement is consistent with the losses of histogram responses.

The FPDE model exhibits superior performance under sparse observation data, a common scenario in real-world applications. For example, under the $2^{-12}$ sampling ratio, the MSE of the converged FPDE model is only 18\% of the baseline. On average, the MSE on the converged FPDE model is 82.1\% less than the baseline model in these 7 sparse cases. However, since the data in this experiment is accurate and the observation data of the first 3 groups are enough in quantity, FPDE models show no improvements. In the magnified part in \fref{fig:sparse result}.a, the first 3 pairs of losses are close to the complete and error-free condition (i.e., case in \fref{fig:3_class}.c). Regarding the conflict, \pref{no conflict} posits an assumption of nearly no conflict between the PDE and data loss. However, in the context of real-world measurements, data are invariably sparse, a scenario in which FPDE models demonstrate enhanced performance.
% xxxIn terms of the conflict, \pref{no conflict} assumes that there is nearly no conflict between PDE and data loss. But in real-world measurements, the data are always sparse, where FPDE models have enhanced performance.

The \fref{fig:sparse result}.b shows the typical outputs in the sparse experiment. The restoration task is similar to the pure equations solving task at the lowest sampling ratio. Using the $2^{-16}$ sampling ratio case as the example, the number of used simulation data points is only 91. To show the sparsity directly, the comparison between the sampled data and full-size data points is shown on the top of \fref{fig:sparse result}.b. The direct NN outputs are shown for the comparison of two models. It is obvious that the PDE solution becomes more trivial (like the example in \fref{fig:failed example}.c). With the filter operation, FPDE model can better capture the characteristic of fluid.

\begin{figure}
    \centering
    \includegraphics[width=\textwidth]{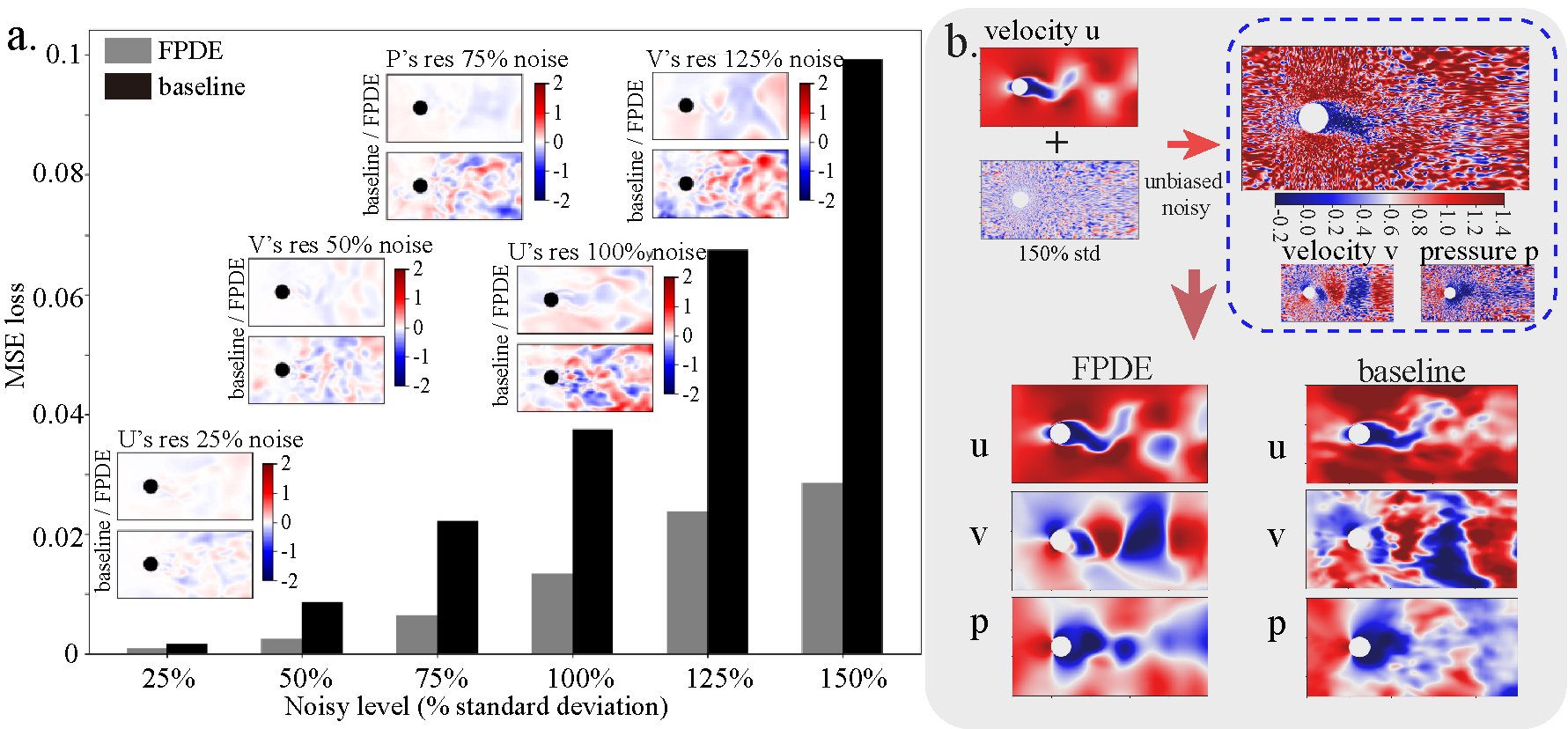}
    \caption{The NN outputs and Mean Squared Error of the FPDE and baseline models under the noisy data condition. (a) The Mean Squared Error and residual plots. (b) The FPDE and PDE outputs under the additive white Gaussian noise with 150\% standard deviation of observation data; the comparison of NN outputs is below.}
    \label{fig:noisy result}
\end{figure}

In the noisy data experiment, the same kind of histogram and residuals are plotted in \fref{fig:noisy result}.a. The conflict with noisy data is shown as \pref{noisy conflict}, which is accurately reflected by the losses in the histogram. Under identical noisy condition, the FPDE model always converged to a lower loss. Meanwhile, the residuals also show FPDE solutions are closer to the ground truth than the baseline model at all noise levels. The \fref{fig:noisy result}.b is a typical example of this noisy group. At the highest noisy level (150\% std, additive white Gaussian noise), the generation of noisy data is shown on the top of \fref{fig:noisy result}.b. Under the high-noise condition, the large-scale features disappear, and periodicity is difficult to reflect. But the FPDE model still shows better anti-noise ability than the baseline model. Only by varying the physics-informed function can the FPDE model significantly reduce the impact of outliers. On average, the MSE on the FPDE model is 62.6\% less than the baseline model in these 7 noisy cases. Under the high-noise data (noisy level $>$ 125\%), this number increases to 72.2\%.

\subsubsection{Experiment results of real-world data}

\begin{figure}
    \centering
    \includegraphics[width=\textwidth]{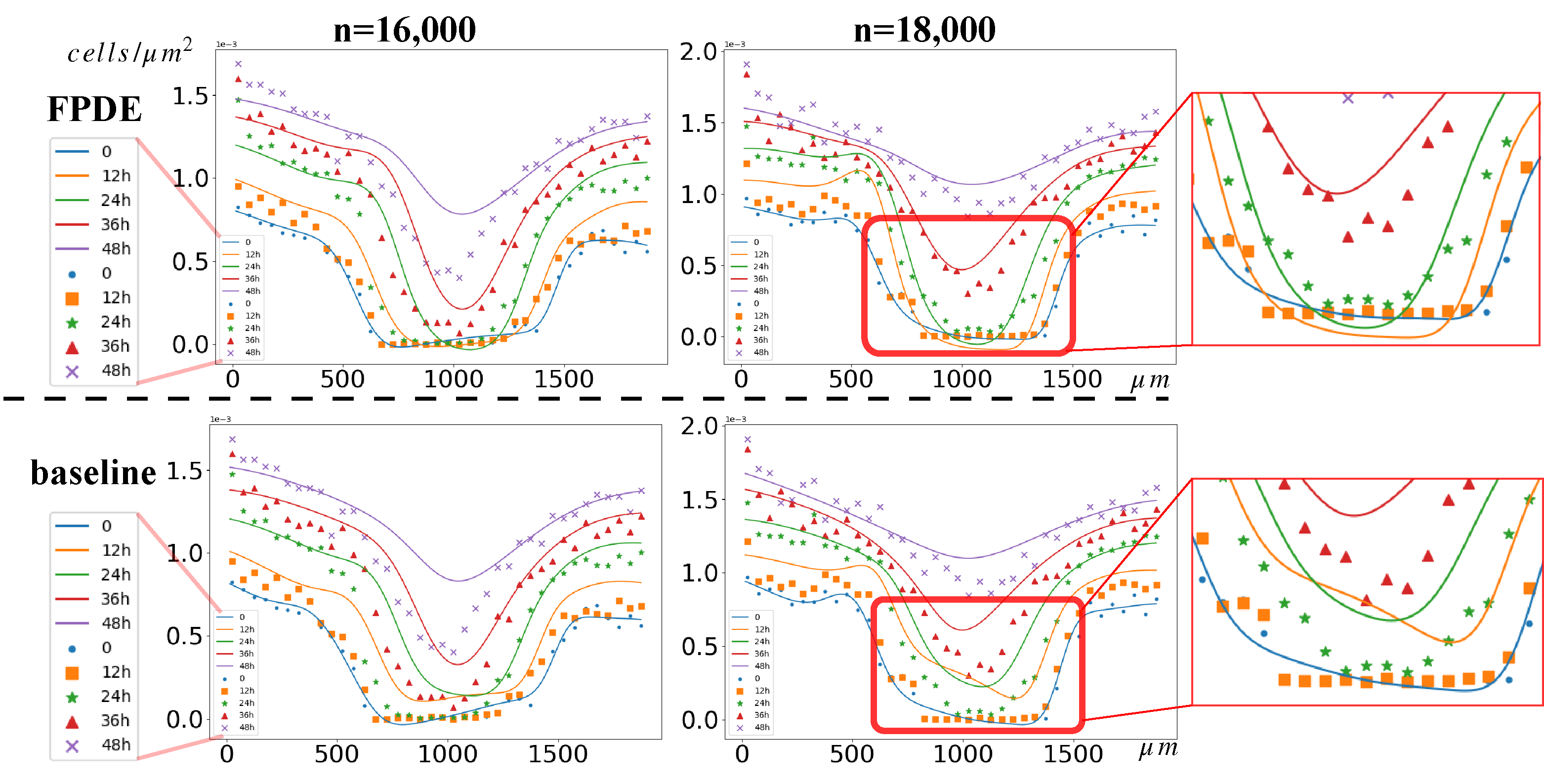}
    \caption{Comparison in the cell migration test data (initial cell number = $[16,000, \ 18,000]$). Inside the red boxes, the difference between two outputs shows FPDE can overcome the influence of outliers better. This characteristic shows that the FPDE method has better modeling ability when faced with real, high-noise measurement data.}
    \label{fig:cm result}
\end{figure}

The data in the cell migration case is sparse and noisy. Additionally, due to fluctuating coefficients, the NN lacks collocation points to learn PDE on the test part. Both experiment results are shown in \fref{fig:cm result} (initial cell number, $n=16,000/18,000$). The predictions indicate the generalized ability of NN to learn the embedding distribution from given data. When confronted with an unknown equation, NN learns the embedding physical process (i.e., the general law of cell migration) through generalization from given equations. In \fref{fig:cm result}, the dots are the measurement data, and the lines are the NN predictions. Quantitatively, the MSE of normalized prediction decreased from 0.157 (baseline) to 0.121 (FPDE), a decrease of 22.9\% when $n=16,000$. When $n=18,000$, this reduction increases to 42.9\%, from $9.89 \times 10^{-2}$ (baseline) to $5.64 \times 10^{-2}$ (FPDE).

In the solutions of the baseline model, it is obvious that the high-noise initial data greatly affects the NN outputs. Outliers cause NN to make predictions that contradict physical laws. Although both models predict worse outcomes over time (e.g., the red lines at t=48h), FPDE model still has more robust results. The unreasonable solution comes due to the inability of the outliers and the physical constraints to effectively correct the prediction well. In the comparison, the FPDE model gave better predictions in the test part. The filter operation mitigates the influence of outliers and makes NN learn the embedding distribution successfully.

\begin{figure}
    \centering
    \includegraphics[width=0.7\textwidth]{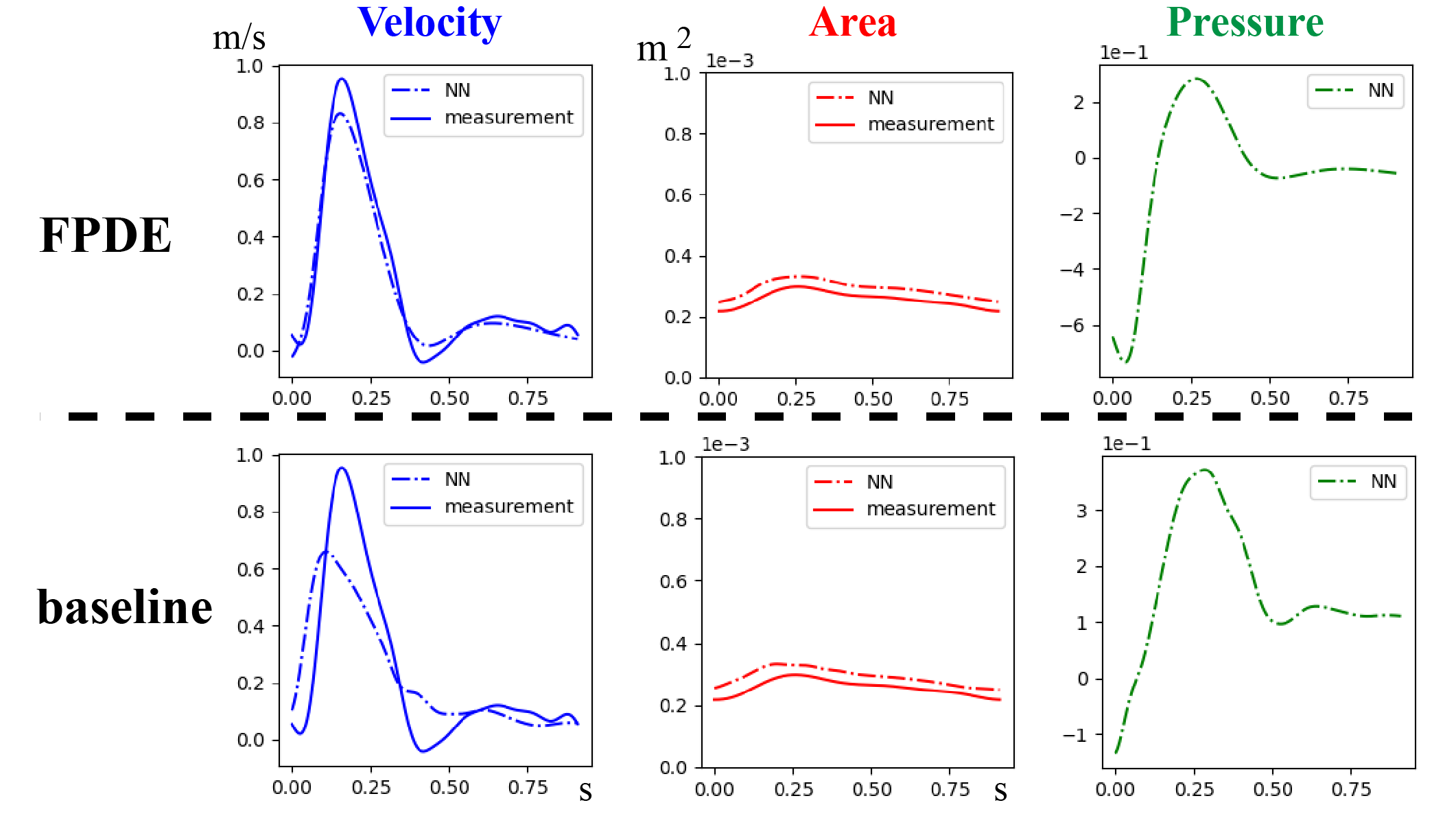}
    \caption{Comparison in the arterial flow test data. The comparison of velocity and area at the test point is used to show the modeling ability of the FPDE and baseline models. Compare with the velocity field in baseline model, FPDE model learns better in large-scale fluctuations.}
    \label{fig:af result}
\end{figure}

The difficulty in modeling the arterial flow data lies in the significant difference between the theoretical vessel model, which is used to derive the equations, and real-world vessels. The NN prediction at point 5 is tested (shown in \fref{fig:af}.b) in the Y-shaped artery with measurement data. The comparison is shown in \fref{fig:af result}. As it demonstrated in the theoretical derivation in section \ref{conflict mechanism}, filter operations do not make NN learn the small-scale fluctuations more finely, but they do make NN model the large-scale information more precisely. On the large-scale of the velocity field, the FPDE model is less inclined to produce trivial results compared to the baseline model. Despite the absence of pressure data in the measurements, NN still gives a prediction of the pressure field through the physical constraints. Since the FPDE model is more accurate in the velocity field, it is intuitive that it has better prediction in the pressure field. Because the equations only constrain the x-direction derivative of pressure, \fref{fig:af result} only shows the NN prediction of pressure individually.

\subsection{Existence of ‘conflict’ in loss analysis}
\begin{figure}
    \centering
    \includegraphics[width=\textwidth]{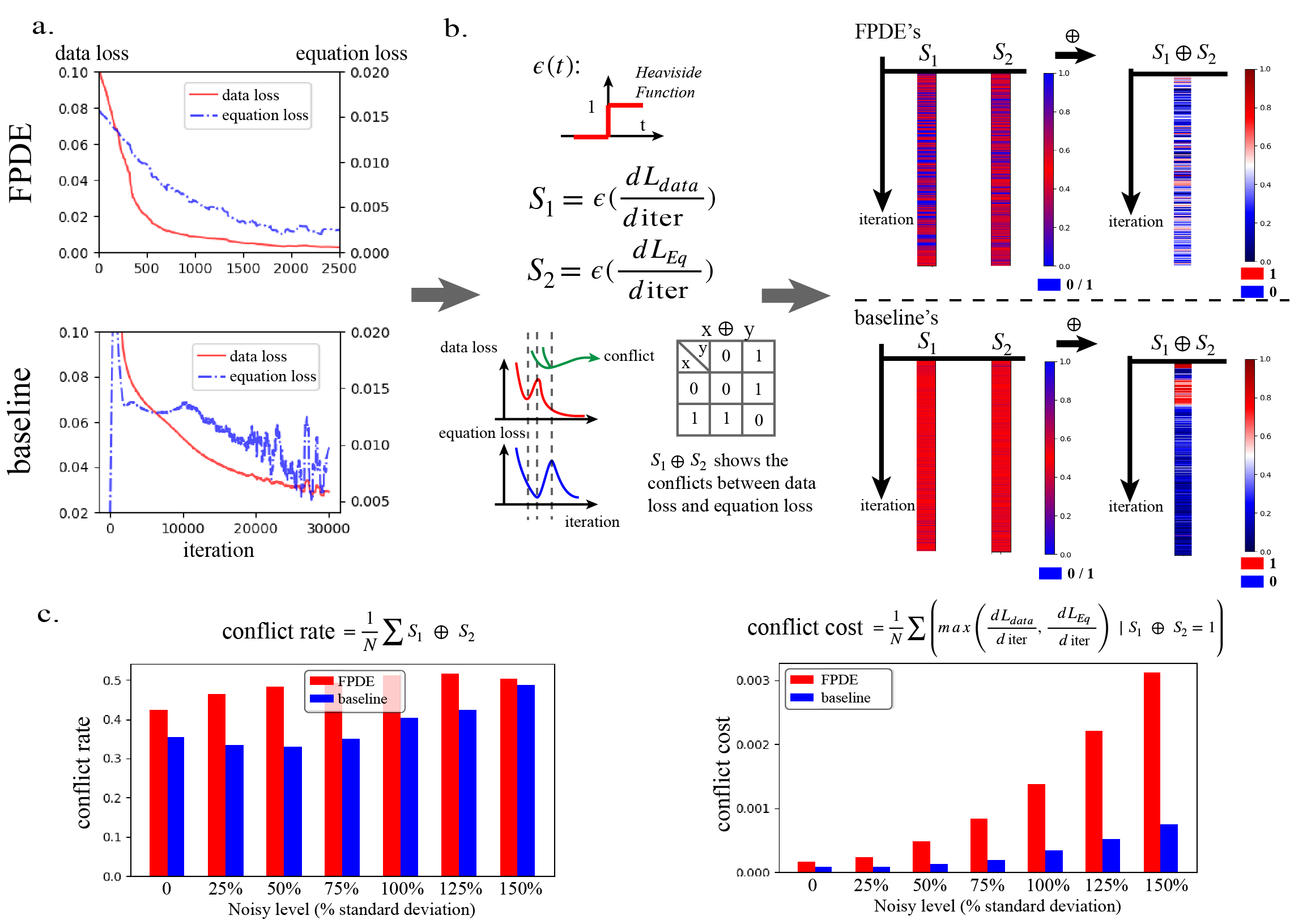}
    \caption{Decoupling achieved by the FPDE method (reflected in training loss). (a) The PDE and data losses of two models until convergence. (b) The S and ‘exclusive OR’ color bars. In order to show the distribution of the derivative sign, the color bar is used in the after-pooling data of the original S. In the S plot, closer to red means more 0-1 oscillating and no effective gradient descent. In the $S_1 \oplus S_2$ plot, blue color indicates no conflict training process. Therefore, the effective descent in the baseline model only happened in the initial part, while the FPDE model realized decoupling throughout the whole training. (c) The definition and statistic result for each group of the conflict rate and conflict cost, which indicates higher conflict tolerance of FPDE than baseline.}
    \label{fig:conflict bar}
\end{figure}

In the motivation part (section 2), we theoretically give the existence of conflict and how the conflict affects the co-optimization process. In our illustration, the conflict can be simply described as the ‘difference between the PDE and data loss optimization direction’. Therefore, we analyze the training losses and find the intuitive data evidence in \fref{fig:conflict bar}.

The comparison between the FPDE model (up) and baseline model (down) is shown in \fref{fig:conflict bar}.a and \fref{fig:conflict bar}.b. Fig.15.a shows the equation loss and data loss of the FPDE and baseline model in the same experiment. In the baseline model, the equation loss is the original PDE loss. Besides, the equation loss is the filtered PDE loss in the FPDE model. To analyze the correlation of the loss decreasing, we defined $S_1$ and $S_2$ as the sign of the equation and data loss derivative. To facilitate the following treatments, the Heaviside function is used to replace the sign operation. Thus, the losses are transferred to the 0/1 lists S, which show the increase or decrease in this iteration. But the 0/1 list is not straightforward, so we change it to blue-red-blue color bar for better demonstration. Specifically, we first average pooling the original data of S and then map the after-pooling data to the blue-red-blue color bar (the color bar on the S left). Therefore, the S bars represent the randomness of loss as they changes by color. Because 0 and 1 represent the direction of loss change, the red color in the middle of the color bar indicates that there is oscillation in this section of the loss. Therefore, when S is in red, the training loss is oscillating, which means the optimization has no effective decrease. Conversely, when the color of the S bar approaches blue, it means the training loss tends to steadily decrease or increase in this segment.

From the color of $S_1$ and $S_2$ bar, it is obvious that the $S_1$ and $S_2$ of baseline model in \fref{fig:conflict bar}.b are almost pure red. Meanwhile, the FPDE model's still have many blue segments. This indicates that the baseline model is more unstable, whereas the FPDE model can optimize effectively.

The last two bars in \fref{fig:conflict bar}.b are the $S_1 \oplus S_2$ bar. $S_1$ and $S_2$ are used to show the directions of losses, thus the $\oplus$ (exclusive OR, xor in short) can be used to show the consistency of $S_1$ and $S_2$. If $S_1$ and $S_2$ are inconsistent ($S_1 \oplus S_2=1$), that means there exist conflict in the co-optimization process. The $S_1 \oplus S_2$ bar is obtained by the exclusive OR solution of the $S_1$ and $S_2$ bars, where 1 means the direction between equation and data loss is different and 0 means the same. After pooling operation, the original data is also visualized with a red-white-blue color bar, where red represents conflicts and blue represents the inverse.

The color of the baseline model in $S_1 \oplus S_2$ shows the conflicts only exist during the initial fast decline segment, and the co-optimization is avoiding the potential conflict and trapping itself in the local optimum in the following segments. Meanwhile, the color in FPDE shows the stochastic conflicts that occured, and the co-optimization does not pay much attention to the potential conflicts, which helps it find the global optimal.
\begin{equation} \label{conflict stat}
	\begin{split}
		conflict\ rate & = \frac{1}{N}\sum S_1 \oplus S_2 \\
		conflict\ cost & = \frac{1}{N}\sum \left(max\left(\frac{d L_{data}}{d L_{train}},\ \frac{d L_{pde}}{d L_{train}} \right)\ |\ S_1 \oplus S_2=1\right) \\
	\end{split}
\end{equation}
 
In \fref{fig:conflict bar}.c, two statistical indicators are proposed to provide additional insights into the observations depicted in \fref{fig:conflict bar}.b. The conflict rate is defined as the frequency of 1 in $S_1 \oplus S_2$ bar, and the conflict cost as the mean derivative of the increasing loss when the conflict happened (shown in \myref{conflict stat}). These indicators are calculated on a group of losses with increasing noise and plot for the general comparison. Though the two indicators exhibiting an upward trend with escalating noise levels, the baseline model still prioritizes conflict avoidance in order to have a lower conflict rate and cost. The higher conflict in the FPDE reflects the decoupling process, which means the correlation between FPDE loss and data loss is decreased.

Overall, the results verify that the optimization process is effectively achieved by diminishing the correlation between the PDE and data loss. In the co-optimization process, the FPDE model can still optimize when facing higher conflict. These findings suggest that the FPDE model demonstrates greater effectiveness in addressing the challenges posed by noise and data sparsity commonly encountered in real-world measurement scenarios.

\section{Conclusions}

Since the data in real-world problems is often insufficient and noisy, it is necessary to improve the learning ability and robustness of the model. In other words, mitigating the reliance of the NN on data quality and quantity is a necessary prerequisite in the modeling of real-world data. One of the difficulties in the training of a physics-informed model is the conflict during the co-optimization. In this paper, we analyzed the causes of the conflict and propose the FPDE method to overcome it. The improvement of the proposed FPDE is verified in three aspects: theoretical derivation, experiments on the simulation data, and the experiments on real-world measurement data. In the comparison between the FPDE and baseline models, the proposed FPDE constraints have the following characteristics:
\begin{itemize}
	\item[$\bullet$] FPDE is a general approach inspired by LES that applies to a variety of models, it is not designed for specific equations or NN architectures. As a surrogate constraint, it has good transferability in most cases.
	\item[$\bullet$] The FPDE method can help NN model the noisy and sparse observation data better. This improvement is important for the practical applications of the physics-informed framework.
	\item[$\bullet$] The FPDE method can better optimize NN when facing mismatches between the data and equations or the equation coefficients are missing.
\end{itemize}

Physics-informed methods have proliferated in recent years, and the FPDE surrogate constraint offers a notable reduction in the model's reliance on data quality and quantity. The proposed FPDE serves as a robust constraint for PINN, exhibiting superior performance in modeling real-world problems with sparse and noisy data, which is highly desirable in practical applications. Moreover, the utilization of FPDE is straightforward, allowing for its application to other models by substituting the original equation constraints with filtered equations during the training process. This article provides an explanation of the mechanism behind the improvement of FPDE from a coupling perspective. The experimental results have shown that the FPDE loss is more prone to escaping local optima, as opposed to being tightly coupled with other losses. This decoupling mechanism not only facilitates easier optimization of FPDE but also leads to improved results, especially in scenarios with sparse and highly noisy data. In future work, additional experiments will be conducted to explore the enhancements of filters in the context of neural networks.

%One of the difficulties in the training of physics-informed model is the ‘Conflict’ in the co-optimization. In chapter 2, we analyzed this conflict, illustrate it in the view of coupling between the PDE and data loss, and raise the filter operation can make decoupling theoretically. Benefit from the optimization feature of NN, the outputs can be obtained first, optimization can be deployed on the filtered outputs. We used the filtered outputs to calculate the residual form of PDE, the FPDE, trained the FPDE model and compared with the original PDE model (PINN) in different quality and quantity of training data. In chapter 4, the comparison and analysis of training shown that the FPDE realized the decoupling in the co-optimization, the gradient descending more smoothly while the PDE model quickly stuck into the local optimum (the S bar in \fref{fig:conflict bar}). In the simulation as well as real measured problem, the FPDE model always give more robust  and reasonable solution than the regular model. But since the cylinder flow problem has strong non-linearity and no stability, the all NN methods still can't learn the periodicity inside. In short, the FPDE constraint, not designed of given equations and given NN architecture, has good mobility and suitable for the complexity of real experiment data. Furthermore, can we improve the structure to make NN learn the hidden periodicity in future? Solving PDE by NN still a long way to go.

\begin{appendices}

\section{Influence of filter in FPDE} \label{appendix filter_type}
\subsection{Filter type}
\begin{figure}
    \centering
    \includegraphics[width=0.5\textwidth]{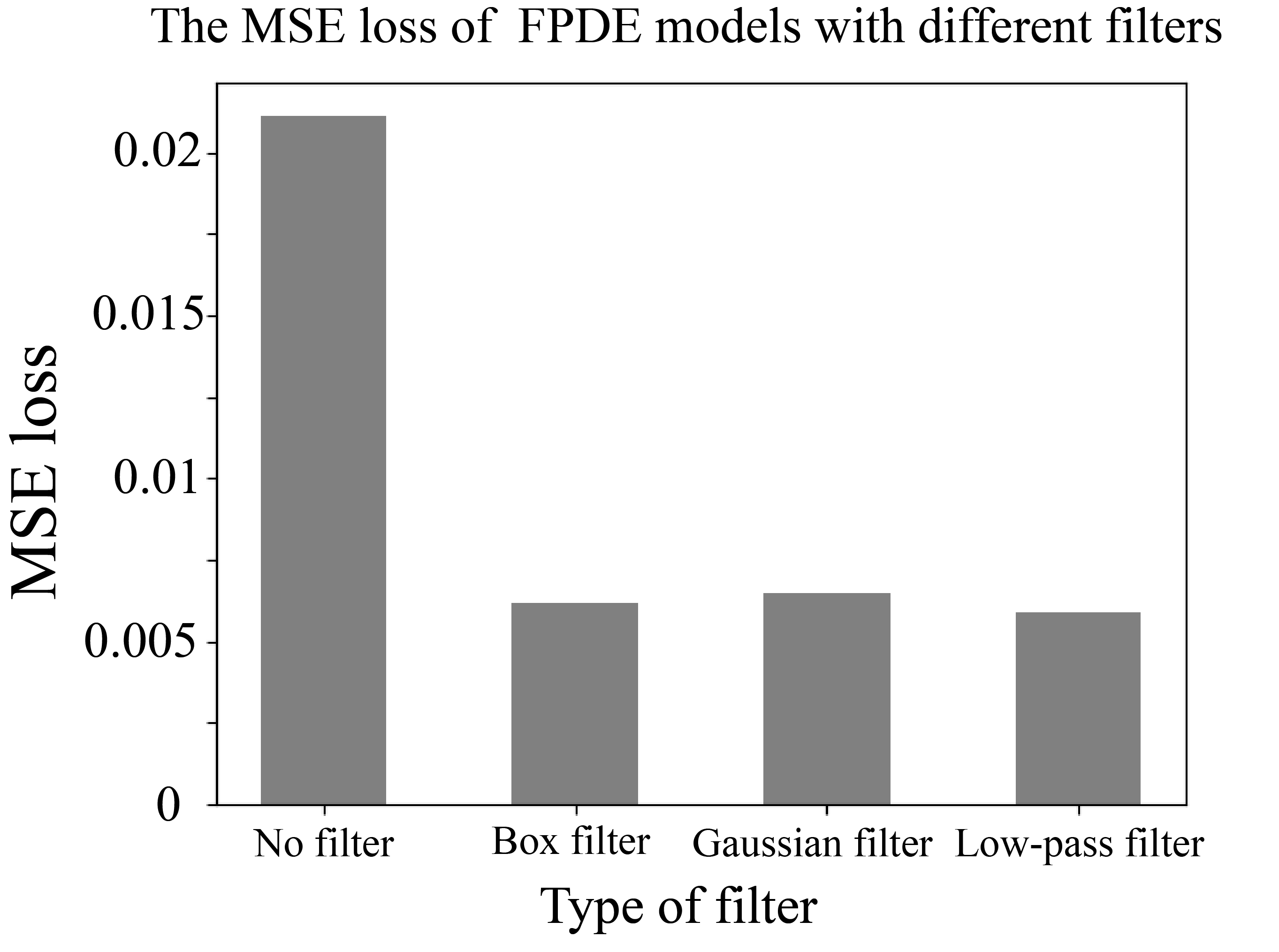}
    \caption{The MSE of the FPDE models with different filters (including ‘no filter’). In this experiment, the chosen classic filters are the box filter, the Gaussian filter, and the low-pass filter. The conventional no-filter model is trained as the control group.}
    \label{fig:type_of_filter}
\end{figure}

In this experiment, the FPDE models with different filters are trained with the same observation data. Thus, the accuracy of results is only determined by the type of filter. The observation data of cylinder flow with a 75\% standard deviation of noise is used. For comparison, the conventional physics-informed model is trained as the control group (i.e., No filter in \fref{fig:type_of_filter}). The mean squared errors (MSE) of the results are shown in the figure below (\fref{fig:type_of_filter}).

The results illustrate that all the given filters in FPDE can improve the training result. Compared to the no-filter group, the improvements of different filters do not exhibit significant discrepancies. It is worth noting that due to the high computational cost of DNS data, only one cylinder flow dataset is utilized in the experiment. This is a typical case used in many studies (e.g., \cite{cheng2021deep}, \cite{rao2020physics}, and \cite{raissi2020hidden}), which facilitates comparisons between different studies. The results in Appendix A are specific to the cylinder flow studied in the experiments and may not be directly applicable to other scenarios.
% It is worth noting that due to the statistics on the results of different fluid cases is not enough, the above conclusions can only be roughly proven on the cylinder flow dataset of this work.
%The result indicates that the FPDE method is robust to different filters and is not sensitive to the selection of filters.

\subsection{Filter size}
\begin{figure}
    \centering
    \includegraphics[width=0.6\textwidth]{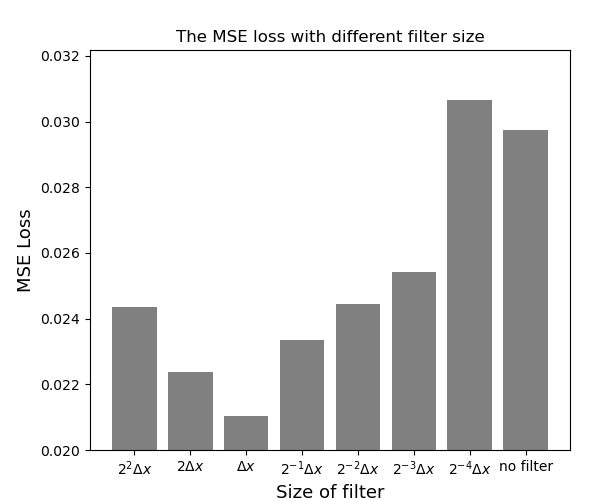}
    \caption{The MSE of the FPDE models with different size Gaussian filters (including ‘no filter’). In this figure, $\Delta x$ is used to stand for the original filter of size 0.1.}
    \label{fig:size_of_filter}
\end{figure}

In the previous experiments, all results were based on the filter size ($\Delta x$) of 0.1. Therefore, this experiment is designed to examine the influence of filter size on FPDE results. Ideally, when the filter size is small enough, the FPDE result should converge to the conventional PINN result. In this experiment, we used training data with a sampling ratio of $2^{-13}$ and $50\% \cdot std$ noise. The chosen filter is the Gaussian filter. With 10 times repeating, the average MSE loss of the converged model on the test data is shown in \fref{fig:size_of_filter}.

The results roughly illustrate that the improvement of FPDE will reach the optimum at an appropriate filter size. A filter size that is too large or too small will increase the MSE loss. It is worth noting that due to the statistics on the results of different fluid cases is not enough, the above conclusions can only be roughly proven on the cylinder flow dataset of this work.

\section{Details of cylinder flow} \label{appendix a}

In this experiment, the selected computation case is the 2-dimensional cylinder flow (shown in 1.1, the background information). The cylinder flow, governed by the N-S equation, is widely studied due to its intricate characteristics. Under the small Reynolds number (Re) condition, the fluid in the cylinder flow case shows the state of Stokes flow or creeping flow \cite{chilcott1988creeping}, which can be simply linearized to the solution of the steady N-S equation. The fluid becomes turbulent as Re increases, and the velocity field shows periodicity. In the vorticity field, the famous Karman vortex street phenomenon appears \cite{wille1960karman}.

The reference IC/BC (initial condition/boundary condition) conditions of the cylinder flow case are from the open resource data in the Supplementary of HFM \cite{raissi2020hidden}, which is generated by OpenFOAM. These conditions are as followes:
\begin{equation}
	\begin{split}
		u(-10, \ y) & = (1, \ 0) \\ 
        u\big(\frac{1}{2}cos\theta, \frac{1}{2}sin\theta\big) & = (0, \ 0)\ , \ \theta \in [0, \ 2\pi] 
	\end{split}
\end{equation}
where zero pressure outflow and periodicity conditions are imposed in the boundary of $[-10, 30] \cdot [-10, 10]$ simulation domain, and the simulation data is selected in $[-2.5, 7.5] \cdot [-2.5, 2.5]$ area.

This simulation data satisfied the N-S equation, which is the ideal observation data in the condition written in \pref{no conflict}. In order to do the restoration in the \fref{fig:3_class}.b, two groups of experiments are designed to showcase the FPDE performances under varying levels of data quality and quantity. Assume that data quality and quantity can be summarized as its noise level and sparse level, thus the simulation data of two groups is shown in Table.1. To apply the control variate method, every case in each group has two parallel models. The only difference between two parallel models is the constraint imposed on the PDE or FPDE loss. The conventional PDE guided model is used as the baseline model for this question, using the same data in both models during each pair of experiments. In order to balance the additional computational cost caused by filter operation in the FPDE model, the batch size of the baseline model is scaled up k times (where k denotes the filter size). Each model is trained for 50,000 iterations and tested in the whole selected domain ($x \in [-2.5, 7.5]$, $y \in [-2.5, 2.5]$ and $t \in [0, 16]$). Thus, the comparison reflects the improvement of the FPDE constraint compared to the conventional PDE constraint.

The training process terminates when the training loss has converged. The final loss function in \myref{general loss} can be written as follows. When training the FPDE model, the variables $y = (u, v, p)$ are filtered initially and are subsequently calculated in the same form as \myref{cf loss}. Each L2 error in the training process is recorded for analysis. The NN parameters are saved every 100 iterations and during iterations where the smallest validation loss is achieved. 

\section{Details of cell migration} \label{appendix b}

In \myref{cm equation}, the constants $D, \lambda \ and \ K$ are always obtained by regression. In the experiments, these constants can be obtained by the ordinary least squares (OLS) method with measurement data. Furthermore, the simplified equations are obtained in the knowledge discovery field in \myref{cms} \cite{chen2021physics}.
\begin{equation} \label{cms}
	\begin{split}
		\frac{\partial C}{\partial t} & = 530.39\frac{\partial^2 C}{\partial x^2} + 0.066C - 46.42C^2, \ n=14,000\\
%		C_t & = 484.74C_{xx} + 0.065C - 43.15C^2 \\
%		C_t & = 636.68C_{xx} + 0.070C - 45.48C^2 \\
		\frac{\partial C}{\partial t} & = 982.26\frac{\partial^2 C}{\partial x^2} + 0.078C - 47.65C^2, \ n=20,000\\
%		& which \ n \ is \ the \ initial \ cells \ number
	\end{split}
\end{equation}
where $n$ is the initial cell number. In this equation, the coefficients $D , \lambda , K$ are decided by initial cell number ($n$), which means the initial density of cell can effect the scratch recover process. 

The measurement data is generated by the cell migration experiment in square petri dishes. Over time, cells migrate and gradually fill in a scratch within the dishes, with cell densities between demarcated blue lines being quantified at 12-hour intervals. Four parallel experiments with different initial cell numbers ($n=14,000$ $/16,000/18,000/20,000$) are measured to show the effect of $n$ on cell density. The example of actual measured data is shown in \fref{fig:cm}.c, which y-axis represents the cell density between adjacent blue lines ($cells/ \mu m^2$) and x-axis is the location ($\mu m$). The density data in $t=0$ is used as the initial data, while subsequent data points ($t=12hours/24hours/36hours/48hours$) are those which the neural network (NN) is tasked with predicting.

\begin{figure}
    \centering
    \includegraphics[width=1\textwidth]{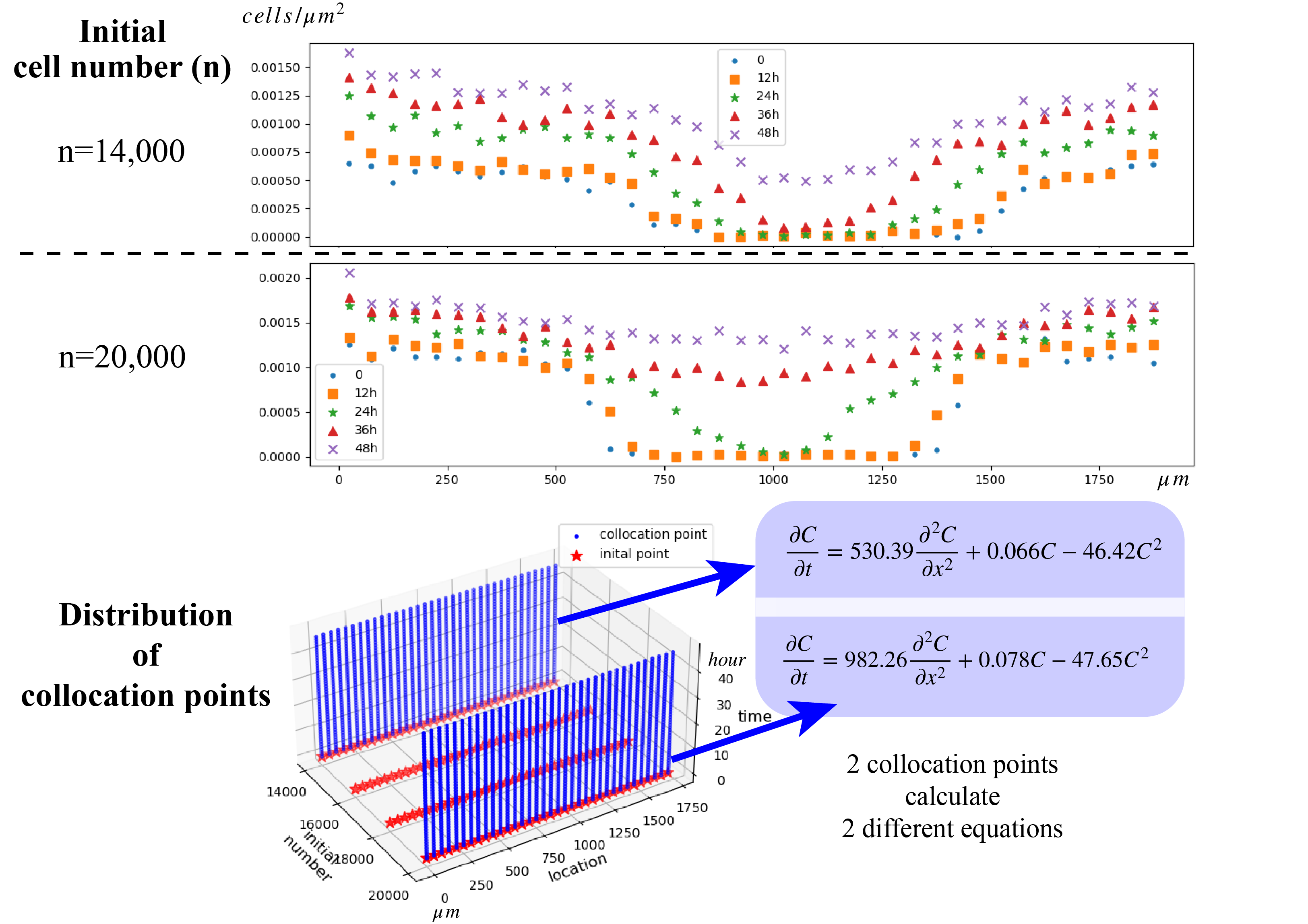}
    \caption{The training data in cell migration modeling. The dataset's initial cell numbers are 14,000 and 20,000 for the NN training. Depending on the initial cell number, different groups fill the scratch at different speeds.}
    \label{fig:cm t_data}
\end{figure}

Obviously, the measurement data is high level of noisy and conducting replicative experiments is prohibitively expensive. Moreover, the data is so sparse in the dimension of n (cell number) that it is impossible to find all coefficients for each n. In terms of the NN training, it is particularly important that the collocation points can't be selected randomly because of the unknown coefficients. Thus it only has the collocation points in $n=14,000 \ and \ 20,000$ flat to calculate \myref{cms}. The task is modeling the distribution of unknown equations ($n \in (14,000, 20,000)$) by the conventional PDE/FPDE method. The training data distribution is shown in \fref{fig:cm t_data}.

\end{appendices}

\section*{Data and materials availability}

All codes used in this manuscript are publicly available on GitHub at \url{https://github.com/Zzzz-Jonathan/FPDE}. Additional data related to this paper may be requested from the authors.

% \bibliographystyle{unsrt}
% \bibliography{FPDE}

\end{document}